\documentclass[journal=nalefd,manuscript=article]{achemso}

\usepackage{bm}
\usepackage{mhchem}
\usepackage{amsmath}
\usepackage{graphicx}
\usepackage[usenames,dvipsnames]{color}
\usepackage[colorlinks,citecolor=magenta]{hyperref}

\usepackage{array}
\usepackage{tabularx}
\usepackage{color}
\usepackage{siunitx}
\usepackage{comment}
\usepackage{xcolor}

\newcommand{\eVA}{\si{\electronvolt\per\angstrom}}
\newcommand{\ket}[1]{\left| #1 \right>}
\newcommand{\matrixel}[3]{\langle #1 | #2 | #3 \rangle}
\mciteErrorOnUnknownfalse

\title{Bonding Character as a Descriptor for Huang-Rhys Factors in Optically Active Defects}
\author{Fatimah Habis}
\affiliation{Department of Physics, University of North Texas, Denton, TX 76207, U.S.A}
\alsoaffiliation{Department of Physical Sciences, Physics Division, Jazan University, P.O. Box. 114 , Jazan 45142, Kingdom of Saudi Arabia}

\author{Yuanxi Wang}
\affiliation{Department of Physics, University of North Texas, Denton, TX 76207, U.S.A}
\email{yuanxi.wang@unt.edu}

\begin{document}

\begin{abstract}
The electron-phonon coupling of a defect – characterized by its Huang–Rhys (HR) factor – is a crucial metric determining its excited-state dynamics, relevant to defect applications as qubits and quantum emitters. However, HR factors remain challenging to calculate from first-principles, complicated by convergence issues in excited-state relaxation and time-consuming phonon calculations. Even when calculated, HR factors lack a rational design principle. Here we show that an orbital-based descriptor can be used to rationalize and efficiently estimate HR factors. Combining this descriptor with a ground-state deformation technique allows circumventing both excited-state relaxation and full phonon calculations. Specifically, our descriptor for HR factors is constructed using bonding-character differences obtained from ground-state density functional theory, measured using crystal orbital Hamilton populations. We demonstrate this descriptor for prototypical hBN defects and the diamond NV– center. This orbital-based descriptor can be potentially used in high-throughput computational screening to identify ideal candidates of spin qubits and SPEs.
\end{abstract}

\maketitle

Point defects in wide-bandgap semiconductors hold great promise for applications as single-photon emitters (SPEs) and spin qubits because they host well-localized mid-gap states \cite{abdi_color_2018,li_identification_2022,jin_photoluminescence_2021,Narayanan_substrate,weber_quantum_2010,alkauskas_first-principles_2014}, similar to isolated atomic levels in trapped ions \cite{zhang_material_2020}. To serve as an ideal SPE or spin qubit, a defect needs to satisfy a long list of criteria \cite{kaul_single_2025, wolfowicz_qubit_2020, weber_quantum_2010, turiansky_rational_2024}, often systematically screened through and catalogued in computational high-throughput studies \cite{bertoldo_quantum_2022, yan_case_2024, cholsuk_hbn_2024, davidsson_adaq_2021}. One key defect property is its electron-phonon coupling, characterized by its Huang–Rhys (HR) factor \cite{huang_theory_1997, li_identification_2022, ivady_ab_2020, tawfik_first-principles_2017, razinkovas_vibrational_2021, jin_photoluminescence_2021, kehayias_infrared_2013}.
For defect-based SPEs, a small HR factor (e.g. $<2$ \cite{GaliMaze2013,  Johansson2011}) is desirable for greater photon indistinguishability. For defect-based qubits, HR factors determine their excited-state dynamics (e.g. non-radiative lifetimes \cite{wu2019carrier,smart_intersystem_2021}), affecting optical initialization, readout, and phonon-induced decoherence \cite{dhara_phonon-induced_2024}.

Despite their importance, HR factors remain hard to calculate and hard to interpret. \textit{Hard to calculate}, in that approximations to excited-state modeling require special care and accurate phonon calculations can be time-consuming at the density functional theory (DFT) level \cite{sharma2025accelerating}. \textit{Hard to interpret}, in that existing HR factor calculation protocols do not deliver rational design rules that can guide experimental efforts towards engineering new defect types. To address the computational challenge, multiple approximation methods are often adopted. An effective one-dimensional (1D) model \cite{jia_first-principles_2017} allows one to circumvent phonon calculations. Excited-state relaxations can be performed with accurate many-body perturbation theory methods~\cite{Sharifzadeh_review,ismail-beigi_excited-state_2003,gao_quasiparticle_2022,tiago_first-principles_2005,kirchhoff_excited-state_2024,del2025revisiting}, but also with the more pragmatic $\Delta$SCF method \cite{gavnholt_general_nodate} at the DFT level, since the it has demonstrated similar accuracies as time-depdent DFT \cite{kowalczyk_assessment_2011} and was recently formalized as an extension of ground-state DFT \cite{yang_foundation_2024}. However, excited-state relaxations using $\Delta$SCF  are still often plagued by convergence issues \cite{kowalczyk_assessment_2011} and level degeneracies.  Even when the 1D model and $\Delta$SCF are applied, the human time involved for validation and convergence checks still puts HR factor calculations (and modeling any phonon-mediated processes) at a computationally challenging level such that they are always relegated to the very last stage of computational screening processes \cite{bertoldo_quantum_2022,sajid_high-throughput_2023,smart_intersystem_2021}, risking prematurely filtering out defects e.g. ones with small HR factor and high defect formation energies that could be engineered to be more stable. Therefore, developing a descriptor to efficiently estimate HR factor would help reposition HR factor calculations to an earlier screening stage that is commensurate with its importance. 

Beyond the computational challenges above, a design challenge remains that there lacks a rational design principle targeting small HR factors: HR factors are often calculated to rank defects performance based on the lowest-energy optical transition without establishing a link to local defect structures or defect orbital characters, which could allow for more targeted defect level engineering. A recent computational high-throughput study of hBN defect property correlations~\cite{cholsuk2025advancing} found HR factors to be largely independent of other defect properties, except for the configuration coordinate (consistent with Ref.~\cite{sajid_high-throughput_2023}); it remains to be answered what local properties are driving excited-state defects to traverse shorter or longer configuration coordinates.
Another recent study \cite{sajid_high-throughput_2023} related higher host material stiffness with lower average HR factors, providing an important guideline for the search of small-HR-factor defects. However, this trend in host-averaged HR factors is overshadowed by even larger variations in defect HR factors \textit{within the same host}. Therefore, identifying a local, defect-specific (instead of host-specific) descriptor is highly desirable to estimate HR factors, ideally one that is able to treat different optical transitions and that avoids the computational challenges stated above. Here, we propose a combination of descriptor design and a deformation technique that avoids both full phonon calculations and excited-state relaxations, all performed based on ground-state DFT. This HR factor estimation approach is numerically efficient and applicable to high-throughput computational screening for identifying ideal defects candidates that serve as SPEs and qubits.

In the following, we will first introduce a descriptor based on bonding characters. Descriptors are obtained from only ground-state Kohn-Sham orbtials, and are used to to estimate excited-state forces, and further calculate HR factors. Initially calibrated by $\Delta$SCF calculations, the descriptor will eventually eliminate the need for $\Delta$SCF. We then introduce a special deformation method to obtain excited-state relaxation displacements. This eliminates both full-phonon calculations and explicit excited-state relaxation.  Taken together, the two approaches require only a single ground-state relaxation. Applying these two approaches, we demonstrate the performance of our descriptor for multiple possible optical transitions in four defect systems -- $\mathrm{V}_{\mathrm{B}}^{-1}, \mathrm{V}_{\mathrm{N}}\mathrm{C}_{\mathrm{B}}$, $\mathrm{V}_{\mathrm{B}}\mathrm{O}_{\mathrm{N}}^{-1}$ in monolayer hexagonal boron nitride (h-BN) \cite{sajid_high-throughput_2023,Sajid_2020,li_identification_2022,abdi_color_2018}, and $\mathrm{NV}^{-1}$ center in bulk diamond. All reference full-phonon calculations employ the embedding approach by Alkauskas et al. \cite{alkauskas_first-principles_2014} to extrapolate interatomic force constants toward the dilute defect limit (see discussion on Fig.~\ref{fig:S-convergence} in the SI). 

\begin{figure}
  \includegraphics[scale=0.4]{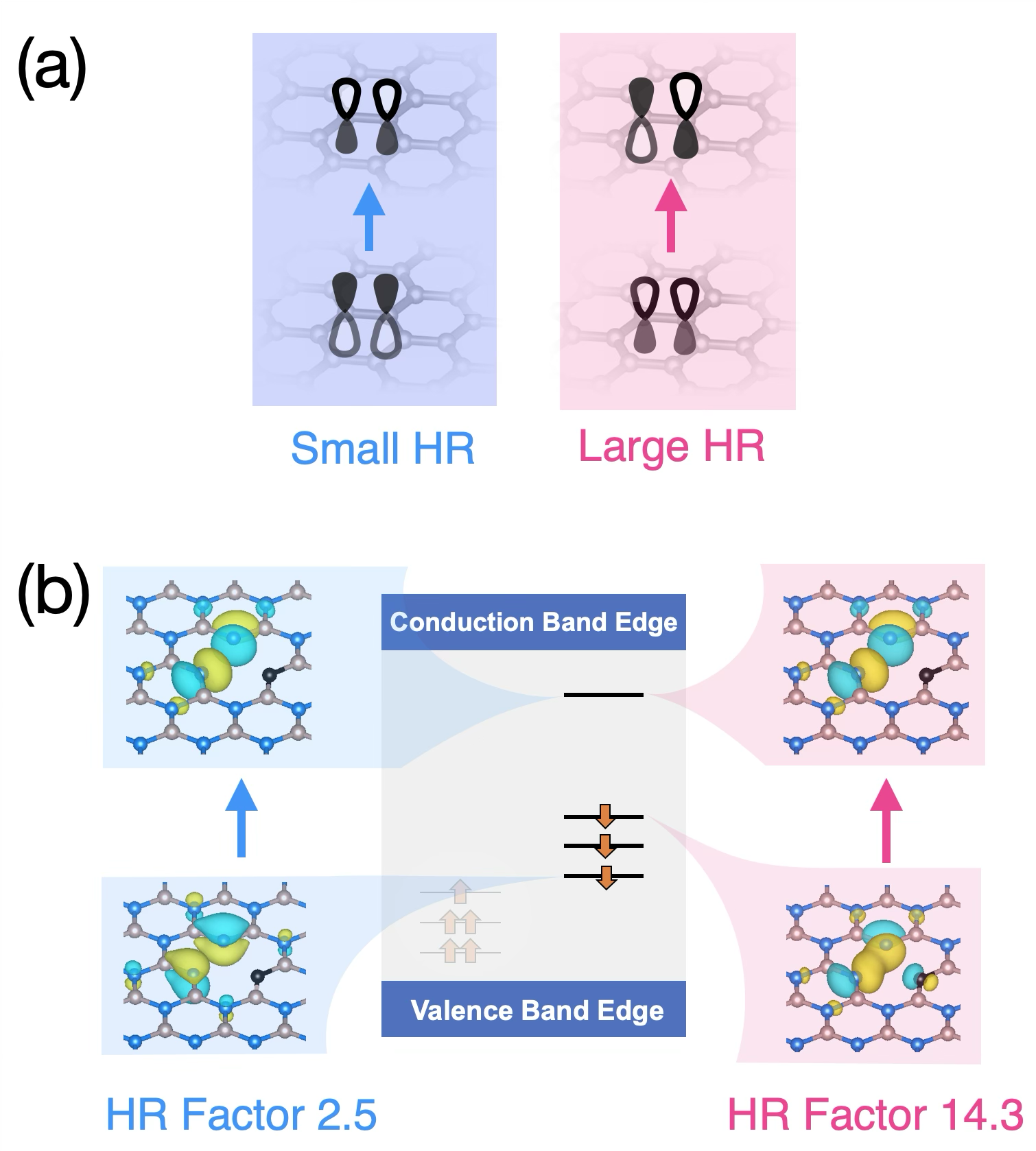}
  \centering
  \caption{(a) HR factors can be associated with the change of bonding character between the initial and final states in a given transition. (b) The wavefunctions and the electronic structure of two possible transitions in the V$_\text{B}$O$_\text{N}$$^{-1}$ defect in monolayer h-BN, showing that a smaller HR factor corresponds to a transition involving two states with similar bonding character, both antibonding.}
  \label{fig:bonding}
\end{figure}

\textbf{Bonding character as a descriptor.}  We first propose that small HR factors are associated with small bonding character changes between the initial (occupied) and final (unoccupied) states involved in an optical transition, i.e. if bonding orbitals remain bonding, or if anti-bonding remains anti-bonding (Fig.~\ref{fig:bonding}a). Here we limit our discussions to single-particle states; the extension to correlated excitations can always be made by including transition weights, i.e. electron-hole amplitudes from time-dependent DFT or solving the Bethe-Salpeter equation (BSE), or equivalently the one-particle transition density matrix more commonly used in excited-state quantum chemistry to apply rules and descriptors originally designed for non-interacting molecular orbitals to correlated many-electron wavefunctions \cite{Kimber2020, Plasser2022, Pokhilko2019}. 
Bonding character has been previously applied in molecular studies to quantify the magnitude of the reorganization energy of molecular excited states \cite{chen_elucidating_2020}. We therefore expect that the same principle would translate to defect systems. As an initial illustrative example, Fig.~\ref{fig:bonding}b shows two optical transitions in the negatively charged V$_\mathrm{B}$O$_\mathrm{N}$$^{-1}$  defect in hBN along with the corresponding two pairs of Kohn-Sham orbital wavefunctions, where the lowest-energy excitation has an HR factor of 14.3 and the next-lowest-energy transition 2.5. The HR factors for the second transition involving a $\pi$ and a $\sigma$ antibonding orbital carries a smaller change in bonding character; the HR factor for this transition is indeed the smaller of the two.

Going beyond this proof-of-principle example, we next derive a quantitative HR factor descriptor based on bonding character. We start from the HR factor itself, a dimensionless quantity representing the average number of phonons of mode $k$ emitted during an electronic transition \cite{huang_theory_1997}.  It is given by $S = \sum_{k} S_{k}$  as a sum over all phonon modes $k$, where partial Huang-Rhys factors are $S_{k} = \omega_{k}^2 q^{2}_{k}/2\hbar \omega_{k}$, with $\omega_k$ being mode frequencies, and $q_k$ being excited-state displacements of atom $i$ projected onto the mode eigenvector ${\vec{e}}_k$ and summed over $i$
\begin{equation}
    \label{eq:q;DeltaR}
    q_k = \sum_{i}\sqrt{M_i} \ (\vec{R}_{i}^{\text{e}}-\vec{R}_{i}^{\text{g}})\cdot \vec{e}_{k;i} 
    = \frac{1}{\omega^{2}_{k}}\sum_{i}\frac{1}{\sqrt{M_i}} \ \vec{F}_{i}^{\text{e}} \cdot \vec{e}_{k;i}
\end{equation}
The second line follows from the harmonic approximation $ \Delta \vec{F} = H \cdot \Delta \vec{R}$,  where $H$ is the Hessian matrix. Phonon frequencies $\omega_k$ are typically taken from phonons at the ground-state equilibrium, where a displaced harmonic oscillator (DHO) approximation is assumed, previously verified for a wide range of defects in solids \cite{jin_photoluminescence_2021,markham_interaction_1959,lax_franckcondon_1952}. ${\vec{F}}_i^\text{e}$ are the forces acting on atom $i$ upon excitation at the equilibrium geometry of the ground state. Therefore a descriptor that captures excited-state forces will serve as a descriptor for HR factors. 

How can bonding characters be associated with excited-state forces? A commonly employed meausure of bonding character is the Crystal Orbital Hamilton Populations (COHP) \cite{dronskowski_crystal_1993}, often useful for qualitatively identifying stabilizing and destabilizing contributions of orbital interactions and comparing phase stabilities \cite{in2se3}, as well as for quantitatively describing bond orders \cite{rohling_correlations_2019}. 
From the total energy of an optically excited state $E_0+E_\text{exc}$ ($E_0$ for ground-state DFT energy and $E_\text{exc}$ for excitation), its derivatives against the 3$N$ ionic positions $R_i$ gives the excited-state forces on each atom $i$,  $F_i^\text{e} = -\partial_{R_i} E_0-\partial_{R_i} E_\text{exc} = -\partial_{R_i} E_\text{exc}$,  with $i=1...3N$, where the first $E_0$ term vanishes at the ground-state equilibrium configuration. To connect towards COHP, we use the simplistic approximation that the derivative of the excitation energy can be described by the derivative of the HOMO and LUMO Kohn-Sham eigenvalue difference \cite{10.1063/1.3526297} $\partial_{R_i} E_\text{exc} \approx  -\partial_{R_i}(\epsilon_c-\epsilon_v)$. In the context of GW-Bethe-Salpeter-Equation approaches to excited-state forces, this approximation neglects excitonic effects, which we do not aspire to include here since our current target is $\Delta$SCF; it also assumes variations in the quasiparticle eigenvalues (and variations in the quasiparticle Hamiltonian) closely follow variations in the Kohn-Sham eigenvalues (and variations in the KS self-consistency potential), as validated in Refs.~\cite{ismail-beigi_excited-state_2003,del2025revisiting} and compared against $\Delta$SCF \cite{ismail-beigi_excited-state_2003}.  From the KS eigenvalue $\epsilon_b$ of a given band $b=c,v$, and following the original COHP formalism~\cite{dronskowski_crystal_1993}, the COHP contribution can be isolated out as off-diagonal terms under a localized-orbital basis set $\ket{\phi_{i\mu}}$, where $i,j,...$ label atoms and $\mu,\nu,...$ label orbitals, i.e. $\epsilon_b = \underset{i,\mu\neq j,\nu}{\sum\sum} H_{i\mu,j\nu} \ \ c_{i\mu,b}^*\ c_{j\nu,b}$, where we collect only off-diagonal terms, $H_{i\mu,j\nu}$ are the Hamiltonian matrix elements $H_{i\mu,j\nu}=\matrixel{\phi_{i\mu}}{\hat{H}}{\phi_{j\nu}}
$, and $c_{i\mu,b}$ are expansion coefficients of band wavefunction into localized orbitals. For small distortions in covalent B-C-N-O systems, assuming that the total energies are mainly affected by covalent bonding, the excited-state forces can be written as a sum of COHPs, 
\begin{equation} 
\begin{split}
    F_i^{\text{e}} &= -\partial_{R_i}(\epsilon_c-\epsilon_v) \\
    &=  -\partial_{R_i} \left[ \sum_{i\mu\neq j\nu}{H_{i\mu,j\nu} \ \ c_{i\mu,c}^*\ c_{j\nu,c}} - (c\leftrightarrow
v)  \right]\\
&\equiv -\partial_{R_i} \left[\sum_{i\mu\neq j\nu} \text{COHP}^c_{i\mu,j\nu}-\text{COHP}^v_{i\mu,j\nu}\right]
\end{split}
\end{equation}
where the second $(c\leftrightarrow
v)$ term is the same as the first except for replacing $c$ subscripts with $v$. Since defect orbitals are localized, and defect bands dispersionless, here we choose to suppress the wavevector dependence and keep the discrete sum over bands, instead of following Ref.~\cite{dronskowski_crystal_1993} to convert into an integral of density-of-states with respect to energy. This is similar to the Molecular Orbital Hamilton Orbital (MOHP) treatment, a discrete analog of COHP \cite{glassey_total_1999}, except that here, our COHP is defined for each state as a parallel to the COHP's original definition as a function of energy, i.e. COHP($E$). Hereafter we will refer to the band-specific COHPs defined here simply as COHP. As usual, negative COHP values indicate bonding interactions, while positive COHP values indicate antibonding interactions. Nonbonding interactions have vanishing COHP values.

In general, the excited-state force can be written as the \textit{difference} in COHP between the ground state and excited state, $ \vec{F}^{\text{e}}_{i}  \approx    -\partial_{{R}_i}\sum_{i\mu\neq j\nu}\Delta\mathrm{COHP}_{i\mu,j\nu}$.  Although converting from KS eigenvalues to COHP values is essentially just a basis transformation from band wavefunctions to localized orbitals, dealing with COHPs has two advantages. (1) They carry physical information about bonding characters allowing direct chemical interpretation, e.g. as previously applied to the strain response of defect emission \cite{dev_fingerprinting}. (2) COHPs and their derivatives always carry the same signs and are highly correlated, allowing further simplifications described as follows. If the ground-to-excited-state change for a given pairwise COHP between atoms $i$ and $j$  ($\Delta $COHP$_{i\mu,j\nu}$) is more negative, it contributes to a net increase in bonding character, and to $\vec F^\text{e}$ being more attractive between atoms $i$ and $j$; same argument applies to more positive $\Delta$COHP$_{i\mu,j\nu}$ contributing to increasing antibonding and $\vec F^\text{e}$ being more repulsive; and zero $\Delta$COHP$_{i\mu,j\nu}$  contributing to no change in bonding character and negligible $\vec F^\text{e}$. We can therefore expect a pairwise $\Delta$COHP value (including sign) to be correlated with its derivative. This expectation can be justified by observing bond-strength-bond-length relations in a wide range of systems, where covalent bond strengths (quantified by integrated COHPs) and bond lengths are highly correlated  -- e.g. for C-C bonds \cite{rohling_correlations_2019}, P-P bonds \cite{zhou_structure_2023}, various bond types in layered transition metal compounds \cite{khazaei_novel_2019}, and across 1520 crystalline compounds \cite{naik_quantum-chemical_2023} -- all related to Pauling's bond-strength-bond-length rule \cite{pauling_atomic_1947}. Since excited-state displacements in B-C-N-O systems ($\sim$0.1~\AA) are much smaller than typical bond lengths, within these small displacement a linear approximation can be taken. Introducing a proportionality factor $A$,  we can write down COHP-estimated forces $\vec{F}_i^\text{COHP}$ as

\begin{equation}
    \label{eq:forces-cohp}
    \vec{F}^{\text{e}}_{i} \approx
    - A \sum_{j\neq i}\sum_{\mu\nu}\Delta\mathrm{COHP}_{i\mu,j\nu} \ \hat{r}_{ij} \equiv \vec{F}_i^\text{COHP}.
\end{equation}
That is, for each atom $i$ and its neighbor $j$, COHP differences $\Delta\mathrm{COHP}_{i\mu,j\nu} $ of the initial state and the final state in a given transition are first summed over all atomic orbital pairs $\mu$ and $\nu$, and assigned along $\hat{r}_{ij}$. Those vectorized COHP differences for each $i,j$ pair are then summed over $j$ and multiplied by $A$  to estimate the total excited-state force acting on atom $i$ (Eq.~\ref{eq:forces-cohp}).  
Using known bond-strength-bond-length relations \cite{rohling_correlations_2019,khazaei_novel_2019,zhou_structure_2023},  the proportionality factor $A$ should be on the order of 2~\AA$^{-1}$. To precisely calibrate $A$, we fit COHP forces with actual excited-state forces from $\Delta$SCF for the collection of hBN and diamond defects mentioned above, yielding $A$ = 3.33~\AA$^{-1}$. Alternatively, $A$ can also be calibrated perturbatively, by monitoring COHP values under finite displacement (e.g. 0.01~\AA), yielding a similar $A$ with 1\% deviation. 

\begin{figure}[h!]
  \includegraphics[scale=0.15]{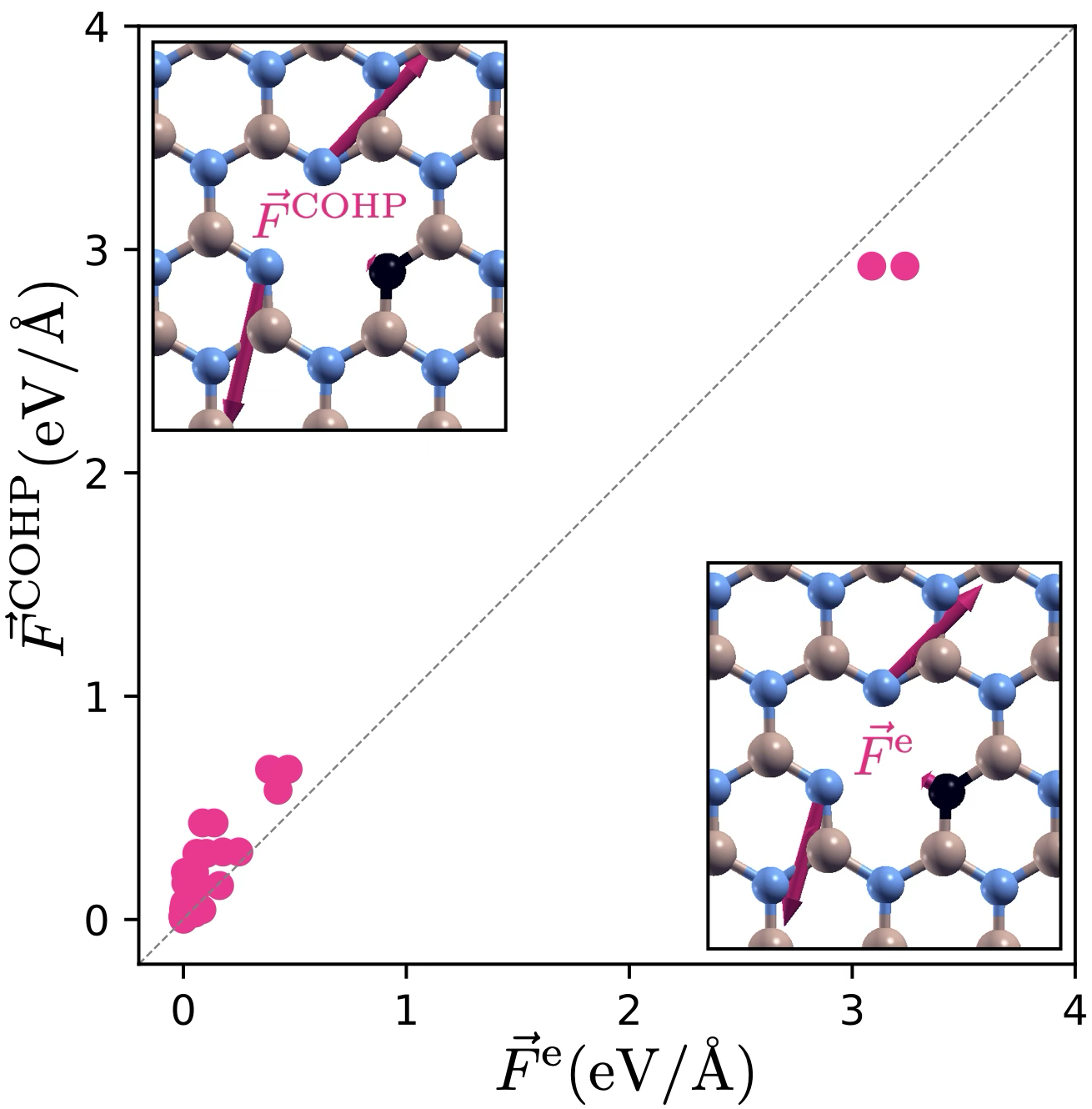}
  \centering
  \caption{For the lowest-energy transition in the V$_\text{B}$O$_\text{N}$$^{-1}$ defect, we compare magnitudes of $\vec{F}^{\text{COHP}}$ versus actual $\Delta$SCF forces $\vec{F}^{\text{e}}$. (Top left inset) Full vector plot of $\vec{F}^{\text{COHP}}$} and (bottom right inset) of $\vec{F}^{\text{e}}$. 
  \label{fig:forces-comparision}
\end{figure}

Fig.~\ref{fig:forces-comparision} shows the magnitudes of all excited-state forces estimated by COHP ($\vec{F}^{\text{COHP}}$) versus actual excited-state forces $\vec{F}^{\text{e}}$ from $\Delta$SCF, for the lowest-energy transition in the defect complex V$_\mathrm{B}$O$_\mathrm{N}$$^{-1}$. The directions of  $\vec{F}^{\text{COHP}}$ (top left inset) also compare well to those of the actual  $\Delta$SCF excited-state forces (bottom right inset). For some defect systems, we observed significant errors in $\vec{F}^{\text{COHP}}$ on selected atoms (see e.g. Fig.~\ref{fig:cohp-forces-outliers} in the SI). However, we show in the following that these force errors, when present, do not propagate onwards to total and partial HR factors. In the SI we explain why partial HR factors are never sensitive to force errors. Thus $\vec{F}^{\text{COHP}}$ is not to be used as a direct replacement of excited-state forces; it serves as a proxy to obtain good estimates of $q_k$, partial HR factors, and total HR factors.

\begin{figure}[t]
  \includegraphics[width=\linewidth]{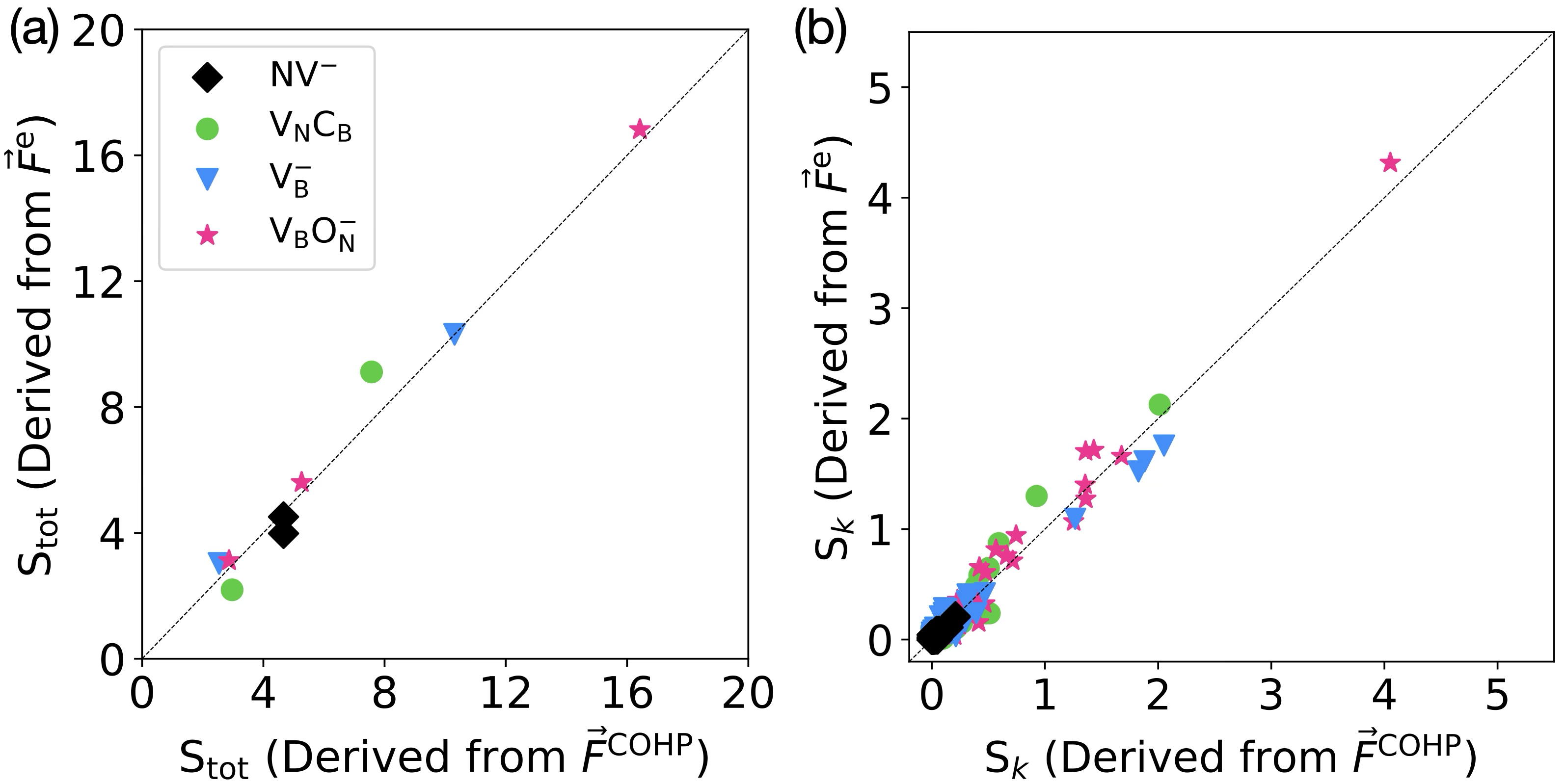}
  \centering
  \caption{(a) Total HR factor obtained from full-phonon calculations using real forces compared to ones obtained from COHP-estimated forces . (b) Comparing partial HR factors for the same set of defects and optical transitions.}
  \label{fig:HR-full-phonon}
\end{figure}

\textbf{Results and comparison with full-phonon calculations. } We now proceed to assess the performance of excited-state forces estimated from COHP against actual ones from $\Delta$SCF based on the prediction accuracy of our final target – the HR factor.  Both forces are applied to Eq.~\ref{eq:q;DeltaR} in combination with full-phonon calculations to calculate normal coordinates $q_k$, and partial and total HR factors. The total HR factor comparison for all defects transitions considered in Fig.~\ref{fig:HR-full-phonon}a shows a good agreement. Specific optical transitions and their corresponding HR factors are reported in Table~\ref{table:HR} in the SI. To assess whether the agreement in HR factors relies on fortuitous error cancellations in partial HR factors $S_{k}$, we perform a stricter assessment comparing $S_{k}$ obtained via COHP-estimated forces versus actual $\Delta$SCF ones for the same set of transitions. An excellent agreement remains in Fig.~\ref{fig:HR-full-phonon}b. It follows that COHP-estimated forces are  good for estimating spectral functions (Eq.~\ref{eq:spectral-function}) as well, as shown in Fig.~\ref{fig:cohp-forces-outliers} in the SI. Quantitative measures of the small estimation errors in $S_k$ are documented in Table~\ref{table:HR} in the SI. 

\begin{figure}[h!]
  \includegraphics[width=0.8\linewidth]{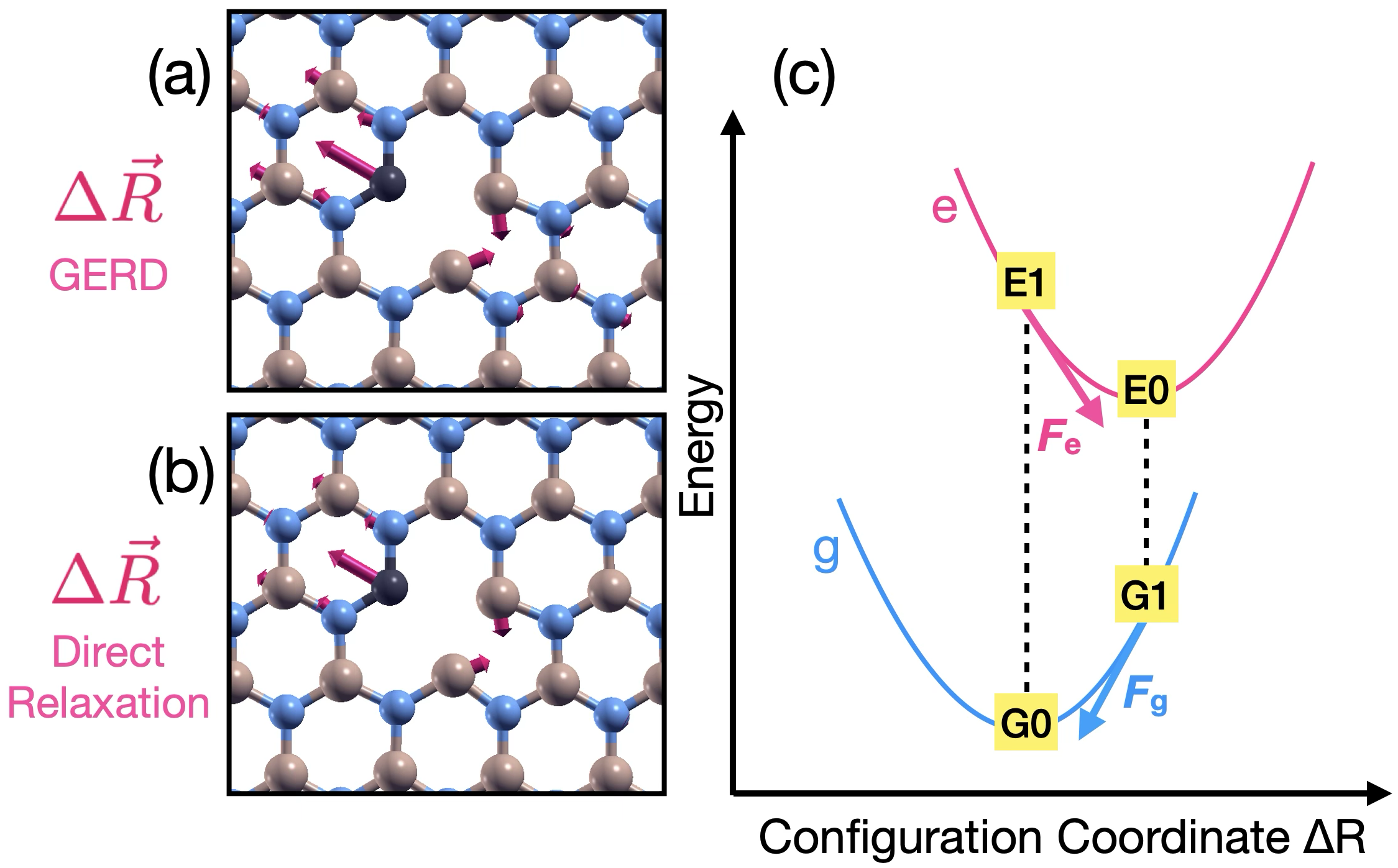}
  \centering
  \caption{(a) Excited-state displacements $\Delta{R}$ obtained via the GERD method and (b) from direct excited-state relaxation with $\Delta$SCF. (c) A schematic of the ground and excited states forces}
  \label{fig:GERD-schematic}
\end{figure}
\textbf{Ground-Excited Reflective Deformation.} Now that COHP-estimated forces have been introduced as an approximation of $\Delta$SCF-based forces, the other numerical challenge remains that an expensive full phonon calculation is still needed to estimate HR factors. Full phonon calculations $can$ be circumvented in the conventional 1D approximation (e.g. as validated for GaN and ZnO \cite{alkauskas_first-principles_2012} and  diamond NV$^{–}$ centers \cite{alkauskas_first-principles_2014}), but the 1D model requires an excited-state \textit{relaxation}, which our proposed one-shot evaluation of COHP-estimated excited-state forces cannot easily adapt to. Can both excited-state relaxations and full phonon calculations be eliminated? To achieve this, we introduce the Ground-Excited Reflective Deformation (GERD) method to determine the excited-state equilibrium without requiring excited-state relaxation. From a typical excitation cycle following the usual 4-point scheme (Fig.~\ref{fig:GERD-schematic}c), we observe that, under the DHO approximation (which is already in place in applying ground-state phonons to Eq.~\ref{eq:q;DeltaR}), the excited-state forces at the ground-state equilibrium [$\vec{F}^{\text{e}}(R^{\text{g}}_\text{eq})$ at Point E1] and the ground-state forces at the excited-state equilibrium [$\vec{F}^{\text{g}}(R^{\text{e}}_\text{eq})$ at Point G1] are equal in magnitude and opposite in directions. The same observation applies to the full potential energy surface in 3$N_\text{atom}$ dimensions. Therefore the search for the excited-state equilibrium configuration is equivalent to the search for the ground-state configuration that yields the initial excited-state forces $\vec{F}^{\text{e}}(R^{\text{g}}_\text{eq})$ with all signs flipped; i.e. we perform a force search in the ground state, not an energy minimization in the excited state. This approach is applicable to any level of modeling excited-state forces (e.g. at the TDDFT or GW-BSE levels), as long as the DHO is assumed.
Practically, this is achieved by imposing the initial excited-state forces $\vec{F}^{\text{e}}(R^{\text{g}}_\text{eq})$ as a set of constant forces to the ground-state equilibrium (Point G0), where the structure will deform and ``relax'' under these imposed forces, until restoring forces reach $\vec{F}^{\text{g}} (R^{\text{e}}_\text{eq})$ at point (Point G1) to balance out the imposed $\vec{F}^{\text{e}}(R^{\text{g}}_\text{eq})$, and the excited-state equilibrium is found. From this point onwards we can follow the standard 1D model, where the total displacement $\Delta R$ between the ground- and excited-state equilibrium is used to calculate the 1D configuration coordinate $Q$ (Eq.~\ref{eq;Q:1dccd} in the SI) and subsequently the effective 1D frequency and HR factor.

\begin{figure}[tb]
  \includegraphics[width=\linewidth]{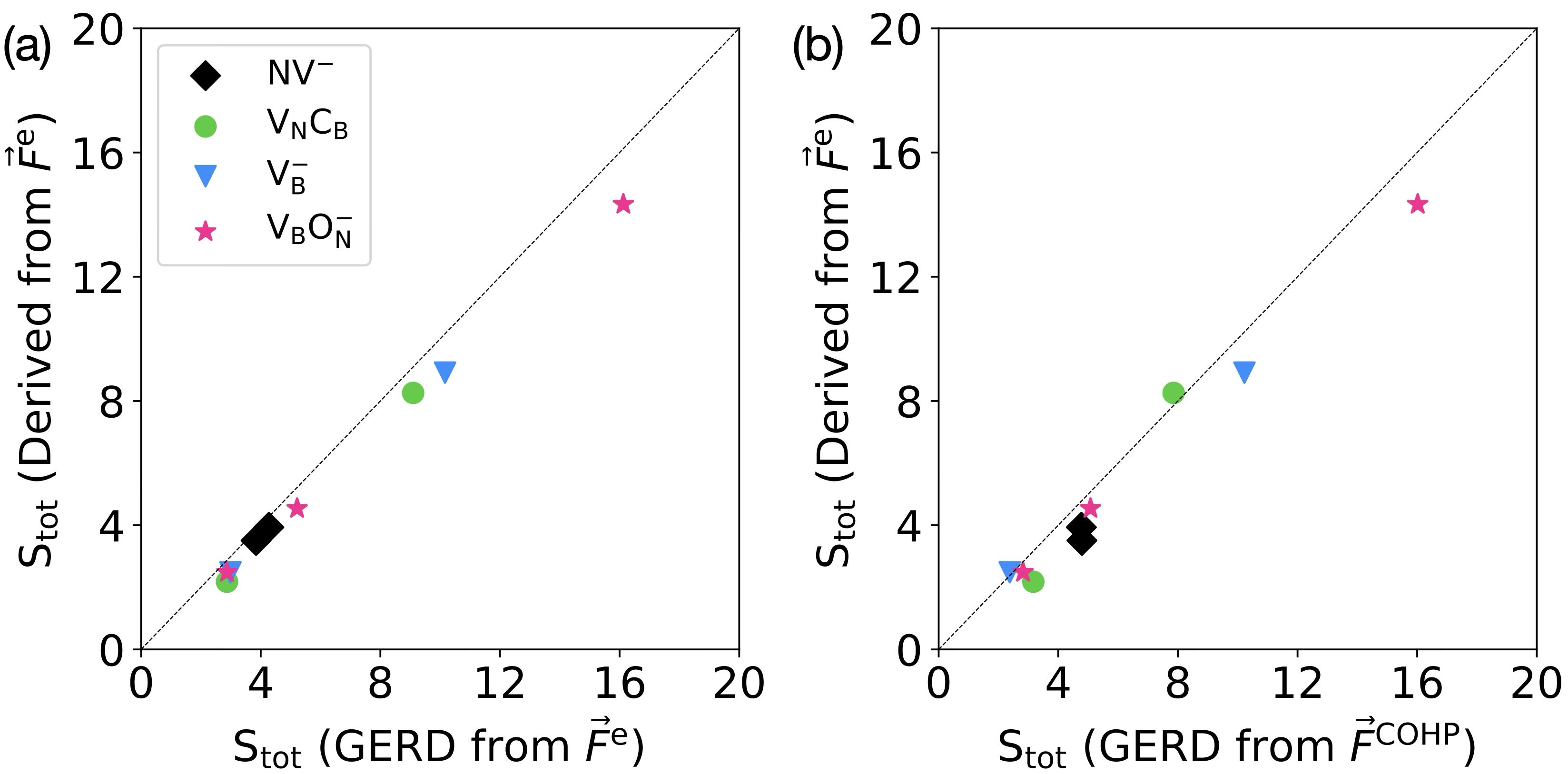}
  \centering
  \caption{Total HR factors obtained from full-phonon calculations using real excited-state forces are used as references (along vertical axis shared across both panels), and compared against HR factors obtained using the GERD method based on (a) real excited-state forces from $\Delta$SCF and (b) COHP-estimated forces. Panel (a) demonstrates the performance of GERD alone; Panel (b) demonstrates the performance of the COHP descriptor combined with GERD. }
  \label{fig:GERD-results}
\end{figure}
As a test for the GERD approach, Fig.~\ref{fig:GERD-schematic}a and b compares the ionic displacements $\Delta R$ obtained from the GERD method using initial forces from $\Delta$SCF versus $\Delta R$ from direct excited-state relaxation with $\Delta$SCF, demonstrating good agreement. As a further quantitative test, in Fig.~\ref{fig:GERD-results}a we compare the total HR factors of all transitions previously considered obtained using $\Delta R$ from GERD (horizontal), versus reference total HR factors obtained from full-phonon calculations (vertical), again showing good agreement. As a final benchmark, we test the combined performance of COHP-estimated forces $\vec{F}^{\text{COHP}}$ and the GERD method, where we impose COHP-estimated initial excited-state forces on the ground-state equilibrium, and proceed with the GERD approach to deform defect structures. The total HR factors obtained this way are compared against reference HR factors from full-phonon calculations in Fig.~\ref{fig:GERD-results}b, showing good agreement across all cases considered. This concludes our validation of both the COHP-based descriptor and the deformation technique introduced above.

To summarize, by developing a COHP descriptor for excited-state forces and combining with a ground-state deformation method, we achieve an efficient approach that circumvents both excited-state relaxation and full phonon calculations. The approach delivers HR factors with good accuracy, benchmarked across a collection of prototypical defects in monolayer h-BN and diamond with HR factors spanning the range from 0 to 20, which encompasses the majority of HR factors for h-BN defects \cite{cholsuk_hbn_2024}. The efficiency of our orbital-based descriptor is expected to be particularly useful for high-throughput defect screening and the rational design of defects with targeted HR factors.

\textbf{Acknowledgments.} Y.W. acknowledges support from NSF DMR-2340733, UNT startup funds, and UNT AMMPI seed funds, as well as computational allocations at the Texas Advanced Computing Center, and NERSC via the Center for Nanophase Materials Sciences user program.

\bibliography{References-abbreviated}

\providecommand{\latin}[1]{#1}
\makeatletter
\providecommand{\doi}
  {\begingroup\let\do\@makeother\dospecials
  \catcode`\{=1 \catcode`\}=2 \doi@aux}
\providecommand{\doi@aux}[1]{\endgroup\texttt{#1}}
\makeatother
\providecommand*\mcitethebibliography{\thebibliography}
\csname @ifundefined\endcsname{endmcitethebibliography}  {\let\endmcitethebibliography\endthebibliography}{}
\begin{mcitethebibliography}{66}
\providecommand*\natexlab[1]{#1}
\providecommand*\mciteSetBstSublistMode[1]{}
\providecommand*\mciteSetBstMaxWidthForm[2]{}
\providecommand*\mciteBstWouldAddEndPuncttrue
  {\def\EndOfBibitem{\unskip.}}
\providecommand*\mciteBstWouldAddEndPunctfalse
  {\let\EndOfBibitem\relax}
\providecommand*\mciteSetBstMidEndSepPunct[3]{}
\providecommand*\mciteSetBstSublistLabelBeginEnd[3]{}
\providecommand*\EndOfBibitem{}
\mciteSetBstSublistMode{f}
\mciteSetBstMaxWidthForm{subitem}{(\alph{mcitesubitemcount})}
\mciteSetBstSublistLabelBeginEnd
  {\mcitemaxwidthsubitemform\space}
  {\relax}
  {\relax}

\bibitem[Abdi \latin{et~al.}(2018)Abdi, Chou, Gali, and Plenio]{abdi_color_2018}
Abdi,~M.; Chou,~J.~P.; Gali,~A.; Plenio,~M.~B. Color {Centers} in {Hexagonal} {Boron} {Nitride} {Monolayers}: {A} {Group} {Theory} and {Ab} {Initio} {Analysis}. \emph{ACS Photonics} \textbf{2018}, \emph{5}, 1967--1976, arXiv: 1709.05414\relax
\mciteBstWouldAddEndPuncttrue
\mciteSetBstMidEndSepPunct{\mcitedefaultmidpunct}
{\mcitedefaultendpunct}{\mcitedefaultseppunct}\relax
\EndOfBibitem
\bibitem[Li and Gali(2022)Li, and Gali]{li_identification_2022}
Li,~S.; Gali,~A. Identification of an {Oxygen} {Defect} in {Hexagonal} {Boron} {Nitride}. \emph{J. Phys. Chem. Lett.} \textbf{2022}, \emph{13}, 9544--9551\relax
\mciteBstWouldAddEndPuncttrue
\mciteSetBstMidEndSepPunct{\mcitedefaultmidpunct}
{\mcitedefaultendpunct}{\mcitedefaultseppunct}\relax
\EndOfBibitem
\bibitem[Jin \latin{et~al.}(2021)Jin, Govoni, Wolfowicz, Sullivan, Heremans, Awschalom, and Galli]{jin_photoluminescence_2021}
Jin,~Y.; Govoni,~M.; Wolfowicz,~G.; Sullivan,~S.~E.; Heremans,~F.~J.; Awschalom,~D.~D.; Galli,~G. Photoluminescence spectra of point defects in semiconductors: {Validation} of first-principles calculations. \emph{Phys. Rev. Materials} \textbf{2021}, \emph{5}, 35--39, arXiv: 2106.08608\relax
\mciteBstWouldAddEndPuncttrue
\mciteSetBstMidEndSepPunct{\mcitedefaultmidpunct}
{\mcitedefaultendpunct}{\mcitedefaultseppunct}\relax
\EndOfBibitem
\bibitem[Narayanan and Dev(2023)Narayanan, and Dev]{Narayanan_substrate}
Narayanan,~S.~K.; Dev,~P. Substrate-Induced Modulation of Quantum Emitter Properties in 2D Hexagonal Boron Nitride: Implications for Defect-Based Single Photon Sources in 2D Layers. \emph{ACS Appl. Nano Mater.} \textbf{2023}, \emph{6}, 3446--3452\relax
\mciteBstWouldAddEndPuncttrue
\mciteSetBstMidEndSepPunct{\mcitedefaultmidpunct}
{\mcitedefaultendpunct}{\mcitedefaultseppunct}\relax
\EndOfBibitem
\bibitem[Weber \latin{et~al.}(2010)Weber, Koehl, Varley, Janotti, Buckley, Van~de Walle, and Awschalom]{weber_quantum_2010}
Weber,~J.~R.; Koehl,~W.~F.; Varley,~J.~B.; Janotti,~A.; Buckley,~B.~B.; Van~de Walle,~C.~G.; Awschalom,~D.~D. Quantum computing with defects. \emph{Proc. Natl. Acad. Sci.} \textbf{2010}, \emph{107}, 8513--8518\relax
\mciteBstWouldAddEndPuncttrue
\mciteSetBstMidEndSepPunct{\mcitedefaultmidpunct}
{\mcitedefaultendpunct}{\mcitedefaultseppunct}\relax
\EndOfBibitem
\bibitem[Alkauskas \latin{et~al.}(2014)Alkauskas, Buckley, Awschalom, and Van De~Walle]{alkauskas_first-principles_2014}
Alkauskas,~A.; Buckley,~B.~B.; Awschalom,~D.~D.; Van De~Walle,~C.~G. First-principles theory of the luminescence lineshape for the triplet transition in diamond {NV} centres. \emph{New J. Phys.} \textbf{2014}, \emph{16}, arXiv: 1405.7313\relax
\mciteBstWouldAddEndPuncttrue
\mciteSetBstMidEndSepPunct{\mcitedefaultmidpunct}
{\mcitedefaultendpunct}{\mcitedefaultseppunct}\relax
\EndOfBibitem
\bibitem[Zhang \latin{et~al.}(2020)Zhang, Cheng, Chou, and Gali]{zhang_material_2020}
Zhang,~G.; Cheng,~Y.; Chou,~J.~P.; Gali,~A. Material platforms for defect qubits and single-photon emitters. \emph{Appl. Phys. Rev.} \textbf{2020}, \emph{7}, 031308, arXiv: 2008.06458\relax
\mciteBstWouldAddEndPuncttrue
\mciteSetBstMidEndSepPunct{\mcitedefaultmidpunct}
{\mcitedefaultendpunct}{\mcitedefaultseppunct}\relax
\EndOfBibitem
\bibitem[Kaul \latin{et~al.}(2025)Kaul, Wang, Li, Li, and Ma]{kaul_single_2025}
Kaul,~A.~B.; Wang,~Y.; Li,~A.-P.; Li,~X.; Ma,~X. Single photon emitters in van der {Waals} solids for quantum photonics: materials, theory and molecular-scale characterization probes. \emph{J. Phys. Appl. Phys.} \textbf{2025}, \emph{58}, 123001\relax
\mciteBstWouldAddEndPuncttrue
\mciteSetBstMidEndSepPunct{\mcitedefaultmidpunct}
{\mcitedefaultendpunct}{\mcitedefaultseppunct}\relax
\EndOfBibitem
\bibitem[Wolfowicz \latin{et~al.}()Wolfowicz, Heremans, Anderson, Kanai, Seo, Gali, Galli, and Awschalom]{wolfowicz_qubit_2020}
Wolfowicz,~G.; Heremans,~F.~J.; Anderson,~C.~P.; Kanai,~S.; Seo,~H.; Gali,~A.; Galli,~G.; Awschalom,~D.~D. \url{http://arxiv.org/abs/2010.16395}, arXiv:2010.16395 [quant-ph]\relax
\mciteBstWouldAddEndPuncttrue
\mciteSetBstMidEndSepPunct{\mcitedefaultmidpunct}
{\mcitedefaultendpunct}{\mcitedefaultseppunct}\relax
\EndOfBibitem
\bibitem[Turiansky \latin{et~al.}(2024)Turiansky, Parto, Moody, and Van~de Walle]{turiansky_rational_2024}
Turiansky,~M.~E.; Parto,~K.; Moody,~G.; Van~de Walle,~C.~G. Rational design of efficient defect-based quantum emitters. \emph{APL Photonics} \textbf{2024}, \emph{9}\relax
\mciteBstWouldAddEndPuncttrue
\mciteSetBstMidEndSepPunct{\mcitedefaultmidpunct}
{\mcitedefaultendpunct}{\mcitedefaultseppunct}\relax
\EndOfBibitem
\bibitem[Bertoldo \latin{et~al.}(2022)Bertoldo, Ali, Manti, and Thygesen]{bertoldo_quantum_2022}
Bertoldo,~F.; Ali,~S.; Manti,~S.; Thygesen,~K.~S. Quantum point defects in {2D} materials - the {QPOD} database. \emph{npj Comput. Mater.} \textbf{2022}, \emph{8}, 1--16\relax
\mciteBstWouldAddEndPuncttrue
\mciteSetBstMidEndSepPunct{\mcitedefaultmidpunct}
{\mcitedefaultendpunct}{\mcitedefaultseppunct}\relax
\EndOfBibitem
\bibitem[Yan \latin{et~al.}(2024)Yan, Kar, Chowdhury, and Bansil]{yan_case_2024}
Yan,~Q.; Kar,~S.; Chowdhury,~S.; Bansil,~A. The {Case} for a {Defect} {Genome} {Initiative}. \emph{Adv. Mater.} \textbf{2024}, \emph{36}, 2303098\relax
\mciteBstWouldAddEndPuncttrue
\mciteSetBstMidEndSepPunct{\mcitedefaultmidpunct}
{\mcitedefaultendpunct}{\mcitedefaultseppunct}\relax
\EndOfBibitem
\bibitem[Cholsuk \latin{et~al.}(2024)Cholsuk, Zand, \c{c}akan, and Vogl]{cholsuk_hbn_2024}
Cholsuk,~C.; Zand,~A.; \c{c}akan,~A.; Vogl,~T. The {hBN} {Defects} {Database}: {A} {Theoretical} {Compilation} of {Color} {Centers} in {Hexagonal} {Boron} {Nitride}. \emph{J. Phys. Chem. C} \textbf{2024}, \emph{128}, 12716--12725\relax
\mciteBstWouldAddEndPuncttrue
\mciteSetBstMidEndSepPunct{\mcitedefaultmidpunct}
{\mcitedefaultendpunct}{\mcitedefaultseppunct}\relax
\EndOfBibitem
\bibitem[Davidsson \latin{et~al.}(2021)Davidsson, Iv\'{a}dy, Armiento, and Abrikosov]{davidsson_adaq_2021}
Davidsson,~J.; Iv\'{a}dy,~V.; Armiento,~R.; Abrikosov,~I.~A. {ADAQ}: {Automatic} workflows for magneto-optical properties of point defects in semiconductors. \emph{Comput. Phys. Commun.} \textbf{2021}, \emph{269}, 108091\relax
\mciteBstWouldAddEndPuncttrue
\mciteSetBstMidEndSepPunct{\mcitedefaultmidpunct}
{\mcitedefaultendpunct}{\mcitedefaultseppunct}\relax
\EndOfBibitem
\bibitem[Huang \latin{et~al.}(1997)Huang, Rhys, and Mott]{huang_theory_1997}
Huang,~K.; Rhys,~A.; Mott,~N.~F. Theory of light absorption and non-radiative transitions in {F}-centres. \emph{Proc. R. Soc. London, Ser. A.} \textbf{1997}, \emph{204}, 406--423\relax
\mciteBstWouldAddEndPuncttrue
\mciteSetBstMidEndSepPunct{\mcitedefaultmidpunct}
{\mcitedefaultendpunct}{\mcitedefaultseppunct}\relax
\EndOfBibitem
\bibitem[Iv\'ady \latin{et~al.}(2020)Iv\'ady, Barcza, Thiering, Li, Hamdi, Chou, Legeza, and Gali]{ivady_ab_2020}
Iv\'ady,~V.; Barcza,~G.; Thiering,~G.; Li,~S.; Hamdi,~H.; Chou,~J.-P.; Legeza,~o.; Gali,~A. Ab initio theory of the negatively charged boron vacancy qubit in hexagonal boron nitride. \emph{npj Comput. Mater.} \textbf{2020}, \emph{6}, 1--6\relax
\mciteBstWouldAddEndPuncttrue
\mciteSetBstMidEndSepPunct{\mcitedefaultmidpunct}
{\mcitedefaultendpunct}{\mcitedefaultseppunct}\relax
\EndOfBibitem
\bibitem[Tawfik \latin{et~al.}(2017)Tawfik, Ali, Fronzi, Kianinia, Tran, Stampfl, Aharonovich, Toth, and Ford]{tawfik_first-principles_2017}
Tawfik,~S.~A.; Ali,~S.; Fronzi,~M.; Kianinia,~M.; Tran,~T.~T.; Stampfl,~C.; Aharonovich,~I.; Toth,~M.; Ford,~M.~J. First-principles investigation of quantum emission from {hBN} defects. \emph{Nanoscale} \textbf{2017}, \emph{9}, 13575--13582\relax
\mciteBstWouldAddEndPuncttrue
\mciteSetBstMidEndSepPunct{\mcitedefaultmidpunct}
{\mcitedefaultendpunct}{\mcitedefaultseppunct}\relax
\EndOfBibitem
\bibitem[Razinkovas \latin{et~al.}(2021)Razinkovas, Doherty, Manson, Van De~Walle, and Alkauskas]{razinkovas_vibrational_2021}
Razinkovas,~L.; Doherty,~M.~W.; Manson,~N.~B.; Van De~Walle,~C.~G.; Alkauskas,~A. Vibrational and vibronic structure of isolated point defects: {The} nitrogen-vacancy center in diamond. \emph{Phys. Rev. B} \textbf{2021}, \emph{104}, 045303\relax
\mciteBstWouldAddEndPuncttrue
\mciteSetBstMidEndSepPunct{\mcitedefaultmidpunct}
{\mcitedefaultendpunct}{\mcitedefaultseppunct}\relax
\EndOfBibitem
\bibitem[Kehayias \latin{et~al.}(2013)Kehayias, Doherty, English, Fischer, Jarmola, Jensen, Leefer, Hemmer, Manson, and Budker]{kehayias_infrared_2013}
Kehayias,~P.; Doherty,~M.~W.; English,~D.; Fischer,~R.; Jarmola,~A.; Jensen,~K.; Leefer,~N.; Hemmer,~P.; Manson,~N.~B.; Budker,~D. Infrared absorption band and vibronic structure of the nitrogen-vacancy center in diamond. \emph{Phys. Rev. B} \textbf{2013}, \emph{88}, 165202\relax
\mciteBstWouldAddEndPuncttrue
\mciteSetBstMidEndSepPunct{\mcitedefaultmidpunct}
{\mcitedefaultendpunct}{\mcitedefaultseppunct}\relax
\EndOfBibitem
\bibitem[Gali and Maze(2013)Gali, and Maze]{GaliMaze2013}
Gali,~A.; Maze,~J.~R. Ab initio study of the split silicon-vacancy defect in diamond: Electronic structure and related properties. \emph{Phys. Rev. B} \textbf{2013}, \emph{88}, 235205\relax
\mciteBstWouldAddEndPuncttrue
\mciteSetBstMidEndSepPunct{\mcitedefaultmidpunct}
{\mcitedefaultendpunct}{\mcitedefaultseppunct}\relax
\EndOfBibitem
\bibitem[D'Haenens-Johansson \latin{et~al.}(2011)D'Haenens-Johansson, Edmonds, Green, Newton, Davies, Martineau, Khan, and Twitchen]{Johansson2011}
D'Haenens-Johansson,~U. F.~S.; Edmonds,~A.~M.; Green,~B.~L.; Newton,~M.~E.; Davies,~G.; Martineau,~P.~M.; Khan,~R. U.~A.; Twitchen,~D.~J. Optical properties of the neutral silicon split-vacancy center in diamond. \emph{Phys. Rev. B} \textbf{2011}, \emph{84}, 245208\relax
\mciteBstWouldAddEndPuncttrue
\mciteSetBstMidEndSepPunct{\mcitedefaultmidpunct}
{\mcitedefaultendpunct}{\mcitedefaultseppunct}\relax
\EndOfBibitem
\bibitem[Wu \latin{et~al.}(2019)Wu, Smart, Xu, and Ping]{wu2019carrier}
Wu,~F.; Smart,~T.~J.; Xu,~J.; Ping,~Y. Carrier recombination mechanism at defects in wide band gap two-dimensional materials from first principles. \emph{Phys. Rev. B} \textbf{2019}, \emph{100}, 081407\relax
\mciteBstWouldAddEndPuncttrue
\mciteSetBstMidEndSepPunct{\mcitedefaultmidpunct}
{\mcitedefaultendpunct}{\mcitedefaultseppunct}\relax
\EndOfBibitem
\bibitem[Smart \latin{et~al.}(2021)Smart, Li, Xu, and Ping]{smart_intersystem_2021}
Smart,~T.~J.; Li,~K.; Xu,~J.; Ping,~Y. Intersystem crossing and exciton–defect coupling of spin defects in hexagonal boron nitride. \emph{npj Comput. Mater.} \textbf{2021}, \emph{7}, 1--8\relax
\mciteBstWouldAddEndPuncttrue
\mciteSetBstMidEndSepPunct{\mcitedefaultmidpunct}
{\mcitedefaultendpunct}{\mcitedefaultseppunct}\relax
\EndOfBibitem
\bibitem[Dhara and Guha(2024)Dhara, and Guha]{dhara_phonon-induced_2024}
Dhara,~P.; Guha,~S. Phonon-induced decoherence in color-center qubits. \emph{Phys. Rev. Research} \textbf{2024}, \emph{6}, 013055\relax
\mciteBstWouldAddEndPuncttrue
\mciteSetBstMidEndSepPunct{\mcitedefaultmidpunct}
{\mcitedefaultendpunct}{\mcitedefaultseppunct}\relax
\EndOfBibitem
\bibitem[Sharma \latin{et~al.}(2025)Sharma, Loew, Wang, Nilsson, Marques, and Thygesen]{sharma2025accelerating}
Sharma,~K.; Loew,~A.; Wang,~H.; Nilsson,~F.~A.; Marques,~M.~A.; Thygesen,~K.~S. Accelerating point defect photo-emission calculations with machine learning interatomic potentials. \emph{arXiv:2505.01403} \textbf{2025}, \relax
\mciteBstWouldAddEndPunctfalse
\mciteSetBstMidEndSepPunct{\mcitedefaultmidpunct}
{}{\mcitedefaultseppunct}\relax
\EndOfBibitem
\bibitem[Jia \latin{et~al.}(2017)Jia, Miglio, Ponc\`{e}, Mikami, and Gonze]{jia_first-principles_2017}
Jia,~Y.; Miglio,~A.; Ponc\`{e},~S.; Mikami,~M.; Gonze,~X. First-principles study of the luminescence of {Eu2}+ -doped phosphors. \emph{Phys. Rev. B} \textbf{2017}, \emph{96}, 1--16, arXiv: 1709.00311\relax
\mciteBstWouldAddEndPuncttrue
\mciteSetBstMidEndSepPunct{\mcitedefaultmidpunct}
{\mcitedefaultendpunct}{\mcitedefaultseppunct}\relax
\EndOfBibitem
\bibitem[Lewis and Sharifzadeh(2021)Lewis, and Sharifzadeh]{Sharifzadeh_review}
Lewis,~D.~K.; Sharifzadeh,~S. Modeling Excited States of Point Defects in Materials from Many-Body Perturbation Theory. \emph{ACS Materials Letters} \textbf{2021}, \emph{3}, 862--874\relax
\mciteBstWouldAddEndPuncttrue
\mciteSetBstMidEndSepPunct{\mcitedefaultmidpunct}
{\mcitedefaultendpunct}{\mcitedefaultseppunct}\relax
\EndOfBibitem
\bibitem[Ismail-Beigi and Louie(2003)Ismail-Beigi, and Louie]{ismail-beigi_excited-state_2003}
Ismail-Beigi,~S.; Louie,~S.~G. Excited-{State} {Forces} within a {First}-{Principles} {Green}'s {Function} {Formalism}. \emph{Phys. Rev. Lett.} \textbf{2003}, \emph{90}, 076401\relax
\mciteBstWouldAddEndPuncttrue
\mciteSetBstMidEndSepPunct{\mcitedefaultmidpunct}
{\mcitedefaultendpunct}{\mcitedefaultseppunct}\relax
\EndOfBibitem
\bibitem[Gao \latin{et~al.}(2022)Gao, Da~Jornada, Ben, Deslippe, Louie, and Chelikowsky]{gao_quasiparticle_2022}
Gao,~W.; Da~Jornada,~F.~H.; Ben,~M.~D.; Deslippe,~J.; Louie,~S.~G.; Chelikowsky,~J.~R. Quasiparticle energies and optical excitations of {3C}-{SiC} divacancy from \${GW}\$ and \${GW}\$ plus {Bethe}-{Salpeter} equation\&nbsp;calculations. \emph{Phys. Rev. Materials} \textbf{2022}, \emph{6}\relax
\mciteBstWouldAddEndPuncttrue
\mciteSetBstMidEndSepPunct{\mcitedefaultmidpunct}
{\mcitedefaultendpunct}{\mcitedefaultseppunct}\relax
\EndOfBibitem
\bibitem[Tiago and Chelikowsky(2005)Tiago, and Chelikowsky]{tiago_first-principles_2005}
Tiago,~M.~L.; Chelikowsky,~J.~R. First-principles {GW}–{BSE} excitations in organic molecules. \emph{Solid State Commun.} \textbf{2005}, \emph{136}, 333--337\relax
\mciteBstWouldAddEndPuncttrue
\mciteSetBstMidEndSepPunct{\mcitedefaultmidpunct}
{\mcitedefaultendpunct}{\mcitedefaultseppunct}\relax
\EndOfBibitem
\bibitem[Kirchhoff \latin{et~al.}(2024)Kirchhoff, Deilmann, and Rohlfing]{kirchhoff_excited-state_2024}
Kirchhoff,~A.; Deilmann,~T.; Rohlfing,~M. Excited-state geometry relaxation of point defects in monolayer hexagonal boron nitride. \emph{Phys. Rev. B} \textbf{2024}, \emph{109}, 085127\relax
\mciteBstWouldAddEndPuncttrue
\mciteSetBstMidEndSepPunct{\mcitedefaultmidpunct}
{\mcitedefaultendpunct}{\mcitedefaultseppunct}\relax
\EndOfBibitem
\bibitem[Del~Grande and Strubbe(2025)Del~Grande, and Strubbe]{del2025revisiting}
Del~Grande,~R.~R.; Strubbe,~D.~A. Revisiting ab-initio excited state forces from many-body Green's function formalism: approximations and benchmark. \emph{arXiv preprint arXiv:2502.05144} \textbf{2025}, \relax
\mciteBstWouldAddEndPunctfalse
\mciteSetBstMidEndSepPunct{\mcitedefaultmidpunct}
{}{\mcitedefaultseppunct}\relax
\EndOfBibitem
\bibitem[Gavnholt \latin{et~al.}(2008)Gavnholt, Olsen, Engelund, and Schi\o{}tz]{gavnholt_general_nodate}
Gavnholt,~J.; Olsen,~T.; Engelund,~M.; Schi\o{}tz,~J. General rights $\Delta$ self-consistent field method to obtain potential energy surfaces of excited molecules on surfaces self-consistent field method to obtain potential energy surfaces of excited molecules on surfaces. \emph{Phys. Rev. B} \textbf{2008}, \emph{78}, 075441\relax
\mciteBstWouldAddEndPuncttrue
\mciteSetBstMidEndSepPunct{\mcitedefaultmidpunct}
{\mcitedefaultendpunct}{\mcitedefaultseppunct}\relax
\EndOfBibitem
\bibitem[Kowalczyk \latin{et~al.}(2011)Kowalczyk, Yost, and Voorhis]{kowalczyk_assessment_2011}
Kowalczyk,~T.; Yost,~S.~R.; Voorhis,~T.~V. Assessment of the $\Delta${SCF} density functional theory approach for electronic excitations in organic dyes. \emph{J. Chem. Phys.} \textbf{2011}, \emph{134}, 054128\relax
\mciteBstWouldAddEndPuncttrue
\mciteSetBstMidEndSepPunct{\mcitedefaultmidpunct}
{\mcitedefaultendpunct}{\mcitedefaultseppunct}\relax
\EndOfBibitem
\bibitem[Yang and Ayers()Yang, and Ayers]{yang_foundation_2024}
Yang,~W.; Ayers,~P.~W. \url{http://arxiv.org/abs/2403.04604}, arXiv:2403.04604 [cond-mat, physics:physics]\relax
\mciteBstWouldAddEndPuncttrue
\mciteSetBstMidEndSepPunct{\mcitedefaultmidpunct}
{\mcitedefaultendpunct}{\mcitedefaultseppunct}\relax
\EndOfBibitem
\bibitem[Ali \latin{et~al.}(2023)Ali, Nilsson, Manti, Bertoldo, Mortensen, and Thygesen]{sajid_high-throughput_2023}
Ali,~S.; Nilsson,~F.~A.; Manti,~S.; Bertoldo,~F.; Mortensen,~J.~J.; Thygesen,~K.~S. High-Throughput Search for Triplet Point Defects with Narrow Emission Lines in 2D Materials. \emph{ACS Nano} \textbf{2023}, \emph{17}, 21105--21115\relax
\mciteBstWouldAddEndPuncttrue
\mciteSetBstMidEndSepPunct{\mcitedefaultmidpunct}
{\mcitedefaultendpunct}{\mcitedefaultseppunct}\relax
\EndOfBibitem
\bibitem[Cholsuk \latin{et~al.}(2025)Cholsuk, Suwanna, and Vogl]{cholsuk2025advancing}
Cholsuk,~C.; Suwanna,~S.; Vogl,~T. Advancing the hBN Defects Database through Photophysical Characterization of Bulk hBN. \emph{arXiv:2507.18093} \textbf{2025}, \relax
\mciteBstWouldAddEndPunctfalse
\mciteSetBstMidEndSepPunct{\mcitedefaultmidpunct}
{}{\mcitedefaultseppunct}\relax
\EndOfBibitem
\bibitem[Sajid and Thygesen(2020)Sajid, and Thygesen]{Sajid_2020}
Sajid,~A.; Thygesen,~K.~S. VNCB defect as source of single photon emission from hexagonal boron nitride. \emph{2D Mater.} \textbf{2020}, \emph{7}, 031007\relax
\mciteBstWouldAddEndPuncttrue
\mciteSetBstMidEndSepPunct{\mcitedefaultmidpunct}
{\mcitedefaultendpunct}{\mcitedefaultseppunct}\relax
\EndOfBibitem
\bibitem[Kimber and Plasser(2020)Kimber, and Plasser]{Kimber2020}
Kimber,~P.; Plasser,~F. Toward an understanding of electronic excitation energies beyond the molecular orbital picture. \emph{Phys. Chem. Chem. Phys.} \textbf{2020}, \emph{22}, 6058--6080\relax
\mciteBstWouldAddEndPuncttrue
\mciteSetBstMidEndSepPunct{\mcitedefaultmidpunct}
{\mcitedefaultendpunct}{\mcitedefaultseppunct}\relax
\EndOfBibitem
\bibitem[Plasser \latin{et~al.}(2022)Plasser, Krylov, and Dreuw]{Plasser2022}
Plasser,~F.; Krylov,~A.~I.; Dreuw,~A. libwfa: Wavefunction analysis tools for excited and open-shell electronic states. \emph{WIREs Computational Molecular Science} \textbf{2022}, \emph{12}, e1595\relax
\mciteBstWouldAddEndPuncttrue
\mciteSetBstMidEndSepPunct{\mcitedefaultmidpunct}
{\mcitedefaultendpunct}{\mcitedefaultseppunct}\relax
\EndOfBibitem
\bibitem[Pokhilko and Krylov(2019)Pokhilko, and Krylov]{Pokhilko2019}
Pokhilko,~P.; Krylov,~A.~I. Quantitative El-Sayed Rules for Many-Body Wave Functions from Spinless Transition Density Matrices. \emph{The Journal of Physical Chemistry Letters} \textbf{2019}, \emph{10}, 4857--4862\relax
\mciteBstWouldAddEndPuncttrue
\mciteSetBstMidEndSepPunct{\mcitedefaultmidpunct}
{\mcitedefaultendpunct}{\mcitedefaultseppunct}\relax
\EndOfBibitem
\bibitem[Chen and Cheng(2020)Chen, and Cheng]{chen_elucidating_2020}
Chen,~W.-C.; Cheng,~Y.-C. Elucidating the {Magnitude} of {Internal} {Reorganization} {Energy} of {Molecular} {Excited} {States} from the {Perspective} of {Transition} {Density}. \emph{J. Phys. Chem. A} \textbf{2020}, \emph{124}, 7644--7657\relax
\mciteBstWouldAddEndPuncttrue
\mciteSetBstMidEndSepPunct{\mcitedefaultmidpunct}
{\mcitedefaultendpunct}{\mcitedefaultseppunct}\relax
\EndOfBibitem
\bibitem[Markham(1959)]{markham_interaction_1959}
Markham,~J.~J. Interaction of {Normal} {Modes} with {Electron} {Traps}. \emph{Rev. Mod. Phys.} \textbf{1959}, \emph{31}, 956--989\relax
\mciteBstWouldAddEndPuncttrue
\mciteSetBstMidEndSepPunct{\mcitedefaultmidpunct}
{\mcitedefaultendpunct}{\mcitedefaultseppunct}\relax
\EndOfBibitem
\bibitem[Lax(1952)]{lax_franckcondon_1952}
Lax,~M. The {Franck}‐{Condon} {Principle} and {Its} {Application} to {Crystals}. \emph{J. Chem. Phys.} \textbf{1952}, \emph{20}, 1752--1760\relax
\mciteBstWouldAddEndPuncttrue
\mciteSetBstMidEndSepPunct{\mcitedefaultmidpunct}
{\mcitedefaultendpunct}{\mcitedefaultseppunct}\relax
\EndOfBibitem
\bibitem[Dronskowski and Bloechl(1993)Dronskowski, and Bloechl]{dronskowski_crystal_1993}
Dronskowski,~R.; Bloechl,~P.~E. Crystal orbital {Hamilton} populations ({COHP}): energy-resolved visualization of chemical bonding in solids based on density-functional calculations. \emph{J. Phys. Chem.} \textbf{1993}, \emph{97}, 8617--8624\relax
\mciteBstWouldAddEndPuncttrue
\mciteSetBstMidEndSepPunct{\mcitedefaultmidpunct}
{\mcitedefaultendpunct}{\mcitedefaultseppunct}\relax
\EndOfBibitem
\bibitem[K{\"u}pers \latin{et~al.}(2018)K{\"u}pers, Konze, Meledin, Mayer, Englert, Wuttig, and Dronskowski]{in2se3}
K{\"u}pers,~M.; Konze,~P.~M.; Meledin,~A.; Mayer,~J.; Englert,~U.; Wuttig,~M.; Dronskowski,~R. Controlled Crystal Growth of Indium Selenide, In2Se3, and the Crystal Structures of $\alpha$-In2Se3. \emph{Inorg. Chem.} \textbf{2018}, \emph{57}, 11775--11781\relax
\mciteBstWouldAddEndPuncttrue
\mciteSetBstMidEndSepPunct{\mcitedefaultmidpunct}
{\mcitedefaultendpunct}{\mcitedefaultseppunct}\relax
\EndOfBibitem
\bibitem[Rohling \latin{et~al.}(2019)Rohling, Tranca, Hensen, and Pidko]{rohling_correlations_2019}
Rohling,~R.~Y.; Tranca,~I.~C.; Hensen,~E. J.~M.; Pidko,~E.~A. Correlations between {Density}-{Based} {Bond} {Orders} and {Orbital}-{Based} {Bond} {Energies} for {Chemical} {Bonding} {Analysis}. \emph{J. Phys. Chem. C} \textbf{2019}, \emph{123}, 2843--2854\relax
\mciteBstWouldAddEndPuncttrue
\mciteSetBstMidEndSepPunct{\mcitedefaultmidpunct}
{\mcitedefaultendpunct}{\mcitedefaultseppunct}\relax
\EndOfBibitem
\bibitem[Fischer \latin{et~al.}(2011)Fischer, Habenicht, Madrid, Duncan, and Prezhdo]{10.1063/1.3526297}
Fischer,~S.~A.; Habenicht,~B.~F.; Madrid,~A.~B.; Duncan,~W.~R.; Prezhdo,~O.~V. Regarding the validity of the time-dependent Kohn–Sham approach for electron-nuclear dynamics via trajectory surface hopping. \emph{J. Chem. Phys.} \textbf{2011}, \emph{134}, 024102\relax
\mciteBstWouldAddEndPuncttrue
\mciteSetBstMidEndSepPunct{\mcitedefaultmidpunct}
{\mcitedefaultendpunct}{\mcitedefaultseppunct}\relax
\EndOfBibitem
\bibitem[Glassey \latin{et~al.}(1999)Glassey, Papoian, and Hoffmann]{glassey_total_1999}
Glassey,~W.~V.; Papoian,~G.~A.; Hoffmann,~R. Total energy partitioning within a one-electron formalism: {A} {Hamilton} population study of surface–{CO} interaction in the c(2×2)-{CO}/ {Ni}(100) chemisorption system. \emph{J. Chem. Phys.} \textbf{1999}, \emph{111}, 893--910\relax
\mciteBstWouldAddEndPuncttrue
\mciteSetBstMidEndSepPunct{\mcitedefaultmidpunct}
{\mcitedefaultendpunct}{\mcitedefaultseppunct}\relax
\EndOfBibitem
\bibitem[Dev(2020)]{dev_fingerprinting}
Dev,~P. Fingerprinting quantum emitters in hexagonal boron nitride using strain. \emph{Phys. Rev. Res.} \textbf{2020}, \emph{2}, 022050\relax
\mciteBstWouldAddEndPuncttrue
\mciteSetBstMidEndSepPunct{\mcitedefaultmidpunct}
{\mcitedefaultendpunct}{\mcitedefaultseppunct}\relax
\EndOfBibitem
\bibitem[Zhou \latin{et~al.}(2023)Zhou, Elliott, and Deringer]{zhou_structure_2023}
Zhou,~Y.; Elliott,~S.~R.; Deringer,~V.~L. Structure and {Bonding} in {Amorphous} {Red} {Phosphorus}. \emph{Angew. Chem. - Int. Ed.} \textbf{2023}, \emph{62}, e202216658\relax
\mciteBstWouldAddEndPuncttrue
\mciteSetBstMidEndSepPunct{\mcitedefaultmidpunct}
{\mcitedefaultendpunct}{\mcitedefaultseppunct}\relax
\EndOfBibitem
\bibitem[Khazaei \latin{et~al.}(2019)Khazaei, Wang, Estili, Ranjbar, Suehara, Arai, Esfarjani, and Yunoki]{khazaei_novel_2019}
Khazaei,~M.; Wang,~J.; Estili,~M.; Ranjbar,~A.; Suehara,~S.; Arai,~M.; Esfarjani,~K.; Yunoki,~S. Novel {MAB} phases and insights into their exfoliation into {2D} {MBenes}. \emph{Nanoscale} \textbf{2019}, \emph{11}, 11305--11314\relax
\mciteBstWouldAddEndPuncttrue
\mciteSetBstMidEndSepPunct{\mcitedefaultmidpunct}
{\mcitedefaultendpunct}{\mcitedefaultseppunct}\relax
\EndOfBibitem
\bibitem[Naik \latin{et~al.}(2023)Naik, Ertural, Dhamrait, Benner, and George]{naik_quantum-chemical_2023}
Naik,~A.~A.; Ertural,~C.; Dhamrait,~N.; Benner,~P.; George,~J. A {Quantum}-{Chemical} {Bonding} {Database} for {Solid}-{State} {Materials}. \emph{Sci. Data} \textbf{2023}, \emph{10}, 610\relax
\mciteBstWouldAddEndPuncttrue
\mciteSetBstMidEndSepPunct{\mcitedefaultmidpunct}
{\mcitedefaultendpunct}{\mcitedefaultseppunct}\relax
\EndOfBibitem
\bibitem[Pauling(1947)]{pauling_atomic_1947}
Pauling,~L. Atomic {Radii} and {Interatomic} {Distances} in {Metals}. \emph{J. Am. Chem. Soc.} \textbf{1947}, \emph{69}, 542--553\relax
\mciteBstWouldAddEndPuncttrue
\mciteSetBstMidEndSepPunct{\mcitedefaultmidpunct}
{\mcitedefaultendpunct}{\mcitedefaultseppunct}\relax
\EndOfBibitem
\bibitem[Alkauskas \latin{et~al.}(2012)Alkauskas, Lyons, Steiauf, and Van~de Walle]{alkauskas_first-principles_2012}
Alkauskas,~A.; Lyons,~J.~L.; Steiauf,~D.; Van~de Walle,~C.~G. First-{Principles} {Calculations} of {Luminescence} {Spectrum} {Line} {Shapes} for {Defects} in {Semiconductors}: {The} {Example} of {GaN} and {ZnO}. \emph{Phys. Rev. Lett.} \textbf{2012}, \emph{109}, 267401\relax
\mciteBstWouldAddEndPuncttrue
\mciteSetBstMidEndSepPunct{\mcitedefaultmidpunct}
{\mcitedefaultendpunct}{\mcitedefaultseppunct}\relax
\EndOfBibitem
\bibitem[Perdew \latin{et~al.}(1996)Perdew, Burke, and Ernzerhof]{perdew_generalized_1996}
Perdew,~J.~P.; Burke,~K.; Ernzerhof,~M. Generalized {Gradient} {Approximation} {Made} {Simple}. \emph{Phys. Rev. Lett.} \textbf{1996}, \emph{77}, 3865--3865\relax
\mciteBstWouldAddEndPuncttrue
\mciteSetBstMidEndSepPunct{\mcitedefaultmidpunct}
{\mcitedefaultendpunct}{\mcitedefaultseppunct}\relax
\EndOfBibitem
\bibitem[Bl\"{o}chl(1994)]{blochl_projector_1994}
Bl\"{o}chl,~P.~E. Projector augmented-wave method. \emph{Phys. Rev. B} \textbf{1994}, \emph{50}, 17953--17953\relax
\mciteBstWouldAddEndPuncttrue
\mciteSetBstMidEndSepPunct{\mcitedefaultmidpunct}
{\mcitedefaultendpunct}{\mcitedefaultseppunct}\relax
\EndOfBibitem
\bibitem[Kresse and Furthm\"{u}ller(1996)Kresse, and Furthm\"{u}ller]{kresse_efficiency_1996}
Kresse,~G.; Furthm\"{u}ller,~J. Efficiency of ab-initio total energy calculations for metals and semiconductors using a plane-wave basis set. \emph{Comput. Mater. Sci.} \textbf{1996}, \emph{6}, 15--50\relax
\mciteBstWouldAddEndPuncttrue
\mciteSetBstMidEndSepPunct{\mcitedefaultmidpunct}
{\mcitedefaultendpunct}{\mcitedefaultseppunct}\relax
\EndOfBibitem
\bibitem[Kresse and Joubert(1999)Kresse, and Joubert]{kresse_ultrasoft_1999}
Kresse,~G.; Joubert,~D. From ultrasoft pseudopotentials to the projector augmented-wave method. \emph{Phys. Rev. B} \textbf{1999}, \emph{59}, 1758--1758\relax
\mciteBstWouldAddEndPuncttrue
\mciteSetBstMidEndSepPunct{\mcitedefaultmidpunct}
{\mcitedefaultendpunct}{\mcitedefaultseppunct}\relax
\EndOfBibitem
\bibitem[Maintz \latin{et~al.}(2016)Maintz, Deringer, Tchougr\'{e}eff, and Dronskowski]{maintz_lobster_2016}
Maintz,~S.; Deringer,~V.~L.; Tchougr\'{e}eff,~A.~L.; Dronskowski,~R. {LOBSTER}: {A} tool to extract chemical bonding from plane-wave based {DFT}. \emph{J. Comput. Chem.} \textbf{2016}, \emph{37}, 1030--1035\relax
\mciteBstWouldAddEndPuncttrue
\mciteSetBstMidEndSepPunct{\mcitedefaultmidpunct}
{\mcitedefaultendpunct}{\mcitedefaultseppunct}\relax
\EndOfBibitem
\bibitem[Togo(2023)]{togo_first-principles_2023}
Togo,~A. First-principles {Phonon} {Calculations} with {Phonopy} and {Phono3py}. \emph{J. Phys. Soc. Japan} \textbf{2023}, \emph{92}, 012001\relax
\mciteBstWouldAddEndPuncttrue
\mciteSetBstMidEndSepPunct{\mcitedefaultmidpunct}
{\mcitedefaultendpunct}{\mcitedefaultseppunct}\relax
\EndOfBibitem
\bibitem[Togo \latin{et~al.}(2023)Togo, Chaput, Tadano, and Tanaka]{togo_implementation_2023}
Togo,~A.; Chaput,~L.; Tadano,~T.; Tanaka,~I. Implementation strategies in phonopy and phono3py. \emph{J. Phys. Condens. Matter} \textbf{2023}, \emph{35}, 353001\relax
\mciteBstWouldAddEndPuncttrue
\mciteSetBstMidEndSepPunct{\mcitedefaultmidpunct}
{\mcitedefaultendpunct}{\mcitedefaultseppunct}\relax
\EndOfBibitem
\bibitem[Alkauskas \latin{et~al.}(2016)Alkauskas, McCluskey, and Van De~Walle]{alkauskas_tutorial_2016}
Alkauskas,~A.; McCluskey,~M.~D.; Van De~Walle,~C.~G. Tutorial: {Defects} in semiconductors—{Combining} experiment and theory. \emph{J. Appl. Phys.} \textbf{2016}, \emph{119}\relax
\mciteBstWouldAddEndPuncttrue
\mciteSetBstMidEndSepPunct{\mcitedefaultmidpunct}
{\mcitedefaultendpunct}{\mcitedefaultseppunct}\relax
\EndOfBibitem
\bibitem[Norambuena \latin{et~al.}(2016)Norambuena, Reyes, Mej\'{\i}a-Lop\'ez, Gali, and Maze]{Norambuena2016}
Norambuena,~A.; Reyes,~S.~A.; Mej\'{\i}a-Lop\'ez,~J.; Gali,~A.; Maze,~J.~R. Microscopic modeling of the effect of phonons on the optical properties of solid-state emitters. \emph{Phys. Rev. B} \textbf{2016}, \emph{94}, 134305\relax
\mciteBstWouldAddEndPuncttrue
\mciteSetBstMidEndSepPunct{\mcitedefaultmidpunct}
{\mcitedefaultendpunct}{\mcitedefaultseppunct}\relax
\EndOfBibitem
\bibitem[Leggett \latin{et~al.}(1987)Leggett, Chakravarty, Dorsey, Fisher, Garg, and Zwerger]{Leggett10987}
Leggett,~A.~J.; Chakravarty,~S.; Dorsey,~A.~T.; Fisher,~M. P.~A.; Garg,~A.; Zwerger,~W. Dynamics of the dissipative two-state system. \emph{Rev. Mod. Phys.} \textbf{1987}, \emph{59}, 1--85\relax
\mciteBstWouldAddEndPuncttrue
\mciteSetBstMidEndSepPunct{\mcitedefaultmidpunct}
{\mcitedefaultendpunct}{\mcitedefaultseppunct}\relax
\EndOfBibitem
\end{mcitethebibliography}

\clearpage

\section*{Supplemental Information}

\subsection{Details on First-Principles Calculations}

We study different optical transitions in three defect systems in monolayer h-BN ($\mathrm{V}_{\mathrm{B}}^{-1}, \mathrm{V}_{\mathrm{N}}\mathrm{C}_{\mathrm{B}}$, $\mathrm{V}_{\mathrm{B}}\mathrm{O}_{\mathrm{N}}^{-1}$, and $\mathrm{NV}^{-1}$ center in bulk diamond. We perform density functional theory calculations using the generalized gradient approximation (GGA-PBE) \cite{perdew_generalized_1996} exchange-correlation functional. 
    Due to the high computational cost of modeling the substantial collection of defects and transitions considered, we refrain from advancing towards using hybrid functionals for force calculations, similar to what is adopted in existing high-throughput studies on defects \cite{bertoldo_quantum_2022, davidsson_adaq_2021}. The choice of using GGA-PBE can also be justified because it only incurs small errors referenced against using hybrid functionals for relaxation or phonons,  smaller than errors in the electronic structure such as zero-phonon lines, which are not the focus of the present study \cite{jin_photoluminescence_2021}.   We use projector-augmented wave pseudopotentials (PAW) \cite{blochl_projector_1994} as implemented by the Vienna Ab-initio Simulation Package VASP \cite{kresse_efficiency_1996,kresse_ultrasoft_1999} to relax the defect structures to their ground states equilibrium. We use an energy cutoff of 500 eV for plane wave expansion and force convergence to 0.002 \eVA. We obtain electronic structures to check for defect levels localized in the band gap that could be used for optical transitions.
COHP values in the ground state are obtained using LOBSTER \cite{maintz_lobster_2016} code. 
 To obtain real excited state forces as our reference, we  use a $7 \times 7 \times 1$ supercell size for all h-BN defects and a $4 \times 4 \times 4$ supercell size for $\mathrm{NV}^{-1}$ center and perform excited states calculations using a single $\Gamma$-point $\Delta$SCF calculations. Phonon frequencies are calculated with the finite displacement approach using PHONOPY software \cite{togo_first-principles_2023, togo_implementation_2023}.  We use a $7 \times 7 \times 1$ supercell and a  $\Gamma$ centered $2\times2\times1$ k-grid for all h-BN phonon calculations, and a $3 \times 3 \times 3$ supercell and one single $\Gamma$-point for $\mathrm{NV}^{-1}$ phonon calculations. We use the dynamical matrix embedding method proposed by Alkauskas et al. \cite{alkauskas_first-principles_2014} to obtain HR factors in $15 \times 15 \times 1$ supercells for h-BN defects and a $10 \times 10 \times 10$ supercell for $\mathrm{NV}^{-1}$. Details for the embedding method can be found later in the SI.

\begin{figure}[ht]
    \includegraphics[width=0.9\linewidth]{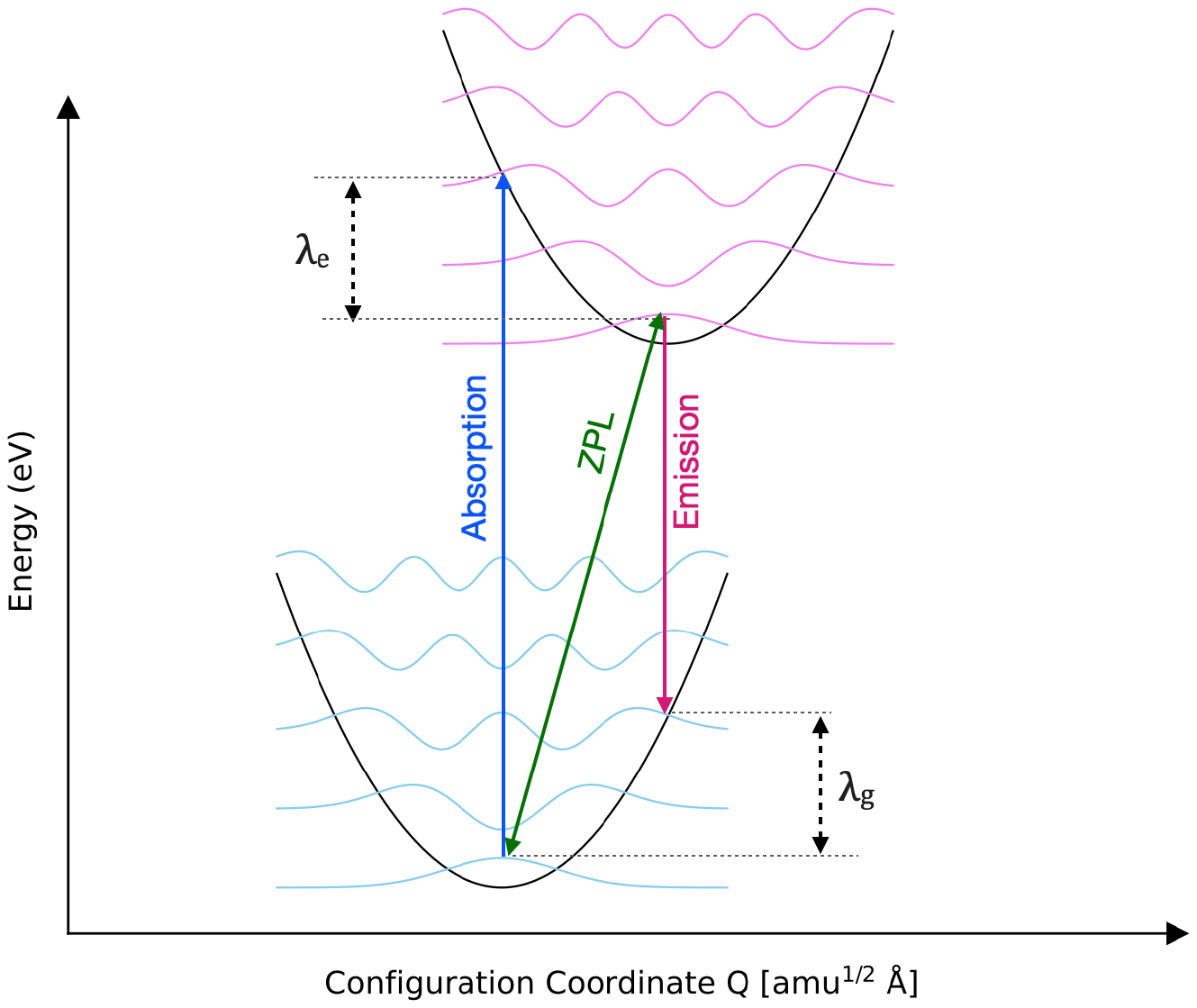}
    \centering
    \caption{The one-dimensional configuration coordinate diagram}
    \label{fig:1dccd}
\end{figure}

\subsection{The One-dimensional Effective Model }

The 1D approximation\cite{alkauskas_tutorial_2016,alkauskas_first-principles_2014, alkauskas_first-principles_2012} shown in Fig.~\ref{fig:1dccd} simplifies all atom displacements upon excitation into one generalized configuration coordinate $Q$, where the change in this configuration coordinate between the ground and excited states equilibrium is represented as: 
\begin{equation}
    \label{eq;Q:1dccd}
    \Delta Q = \sum_{i}{\sqrt{M_i}} \ ({\vec{R}_i^{\text{e}} - \vec{R}_i^{\text{g}}})
\end{equation}
Where $\vec{R}_i^{\text{g;e}}$ are the cartesian atomic positions in the ground and excited states equilibrium.  In this approximation, we consider one effective vibrational mode, with one frequency: 
\begin{equation}
    {\Omega^2_{\text{e}}} = \frac{2\lambda_{\text{e}}}{\Delta Q^2}
\end{equation}
\begin{equation}
    {\Omega^2_{\text{g}}} = \frac{2\lambda_{\text{g}}}{\Delta Q^2}
\end{equation}
Where $\lambda$ is the relaxation energy. The equal-mode approximation within the 1D approximation assumes that these two frequencies are the same effective frequency $\Omega_\text{eff} $.

The total HR factor can be obtained in this 1D model:
\begin{equation}
    S_{\text{tot}} = \frac{\lambda_{\text{e;g}}}{\hbar \Omega_\text{eff}}
\end{equation}

\subsection{Force constants Extrapolation}
\begin{figure}[h!]
  \includegraphics[scale=0.18]{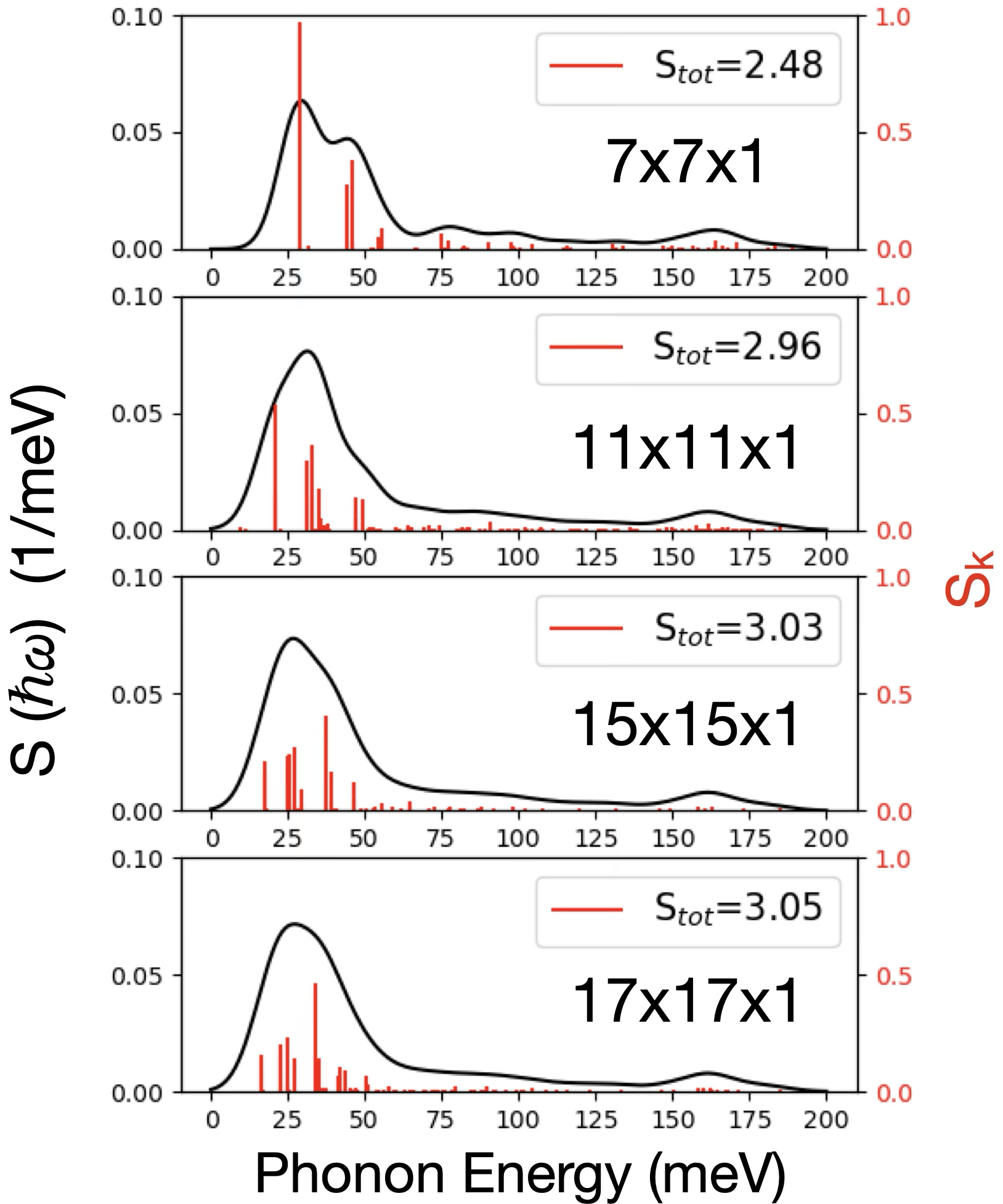}
  \centering
  \caption{Spectral function convergence for the first transition in $\mathrm{V}_{\mathrm{B}}^{-1}$}
  \label{fig:S-convergence}
\end{figure}

To obtain converged partial HR factors and spectral functions 
\begin{equation}
    S(\hbar\omega) = \sum_{k}{S_{k}\delta(\hbar\omega-\hbar\omega_{k})}
    \label{eq:spectral-function}
\end{equation}
in the dilute defect limit, we employ the force constants embedding method  \cite{alkauskas_first-principles_2014}, where the excited-state forces of a smaller supercell is embedded in the large supercell such that all excited-state forces that are far away from the defects are set to zero. The force constants matrix for a large supercell is constructed by extrapolating the elements from a dynamical matrix of a small supercell and applying two cutoffs, $R_1$ and $R_2$. If two atoms in a pair are further away from each other by a distance larger than $R_1$, we set the force constant element between them to zero. If one of the atoms is within a radius less than $R_2$ from the defect center, we take the force constant matrix element from the small defect supercell. For all other pairs, the elements are taken from the force constants matrix of the bulk. 
Previous studies showed that for $\mathrm{NV}^{-1}$, excited-state forces approached zero further than 7 \si{\angstrom} away from the vacancy center \cite{alkauskas_first-principles_2014, jin_photoluminescence_2021}. Thus, We used excited-state forces computed for a $4\times4\times4$ supercell.  To construct the force constants matrix, vibrational frequencies and modes were calculated for a $3\times3\times3$ supercell and we used $R_1$ = 5 \si{\angstrom} and $R_2$ = 5 \si{\angstrom}. These cutoffs are the same as those tested in Ref.~\cite{jin_photoluminescence_2021}. 
For defects in hBN, our calculations show that exited-state forces also decay to zero when are more than 7 \si{\angstrom} away from the defect center. We calculated the excited-state forces and dynamical matrix elements for a $7\times7\times1$ supercell. To choose our cutoffs, we calculated the spectral function from full-phonon for a $9\times9\times1$ supercell for the first transition in $\text{V}_{\text{B}}^{-1}$ to take as a reference. We then tested a range of different values for $R_1$ and $R_2$ and compared the spectral function and total HR factor obtained by embedding a $7\times7\times1$ supercell into a $9\times9\times1$ supercell to our reference. Our tests show that the choice of $R_1$ = 6.28 \si{\angstrom} and $R_2$ = 8.79 \si{\angstrom} gave a converged spectral function and total HR factor. To test the convergence across supercell sizes, we applied those cutoffs for the same first transition in $V_{B}^{-1}$ and embedded the $7\times7\times1$ supercell dynamical matrix into a range of larger supercell sizes. Fig.~\ref{fig:S-convergence} shows that the spectral function and total HR factor both converge at a $15\times15\times1$ supercell. Because all our studied defects in hBN involve a centered vacancy and a next-neighbor  substitutional defect, we apply those same cutoffs values and a supercell size of $15\times15\times1$ to calculate all HR factors for hBN defects using the embedding method. Total HR factors and root mean square errors in partial HR factors for all defect transitions are listed in Table \ref{table:HR}.

\begin{table*}
\centering
\renewcommand{\arraystretch}{1.5} % Adjusts the row height (1.5 is a multiplier)
\begin{tabular}{|>{\centering\arraybackslash}m{2cm}|>{\centering\arraybackslash}m{2cm}|>{\centering\arraybackslash}m{4cm}|>{\centering\arraybackslash}m{2cm}|>{\centering\arraybackslash}m{2cm}|>{\centering\arraybackslash}m{2cm}|}

\hline
%\multicolumn{5}{|c|}{Total HR Factors $S$} \\

\hline
\textbf{Defect} & Spin channel & Transition & $S_{\vec{F}^{\text{e}}}$ & $S_{\vec{F}^{\text{COHP}}}$ & RMSE ($S_{k}$) \\

\hline
{$\mathrm{NV}^{-1}$} 
& 2 & (HOMO $\rightarrow$ LUMO) & 4.51 & 4.67 & 0.001 \\
& 2 & (HOMO $\rightarrow$ LUMO + 1) & 3.98 & 4.67 & 0.001 \\

\hline
{$\mathrm{V}_{\mathrm{N}}\mathrm{C}_{\mathrm{B}}$} 
& 1 & (HOMO $\rightarrow$ LUMO) & 2.19 & 2.97 & 0.01 \\
& 1 & (HOMO $-$ 1 $\rightarrow$ LUMO) & 9.11 & 7.57 & 0.018 \\

\hline
{${\mathrm{V}_{\mathrm{B}}\mathrm{O}_{\mathrm{N}}}^{-1}$} 
& 2 & (HOMO $-$ 2 $\rightarrow$ LUMO) & 3.12 & 2.87 & 0.008 \\
& 2 & (HOMO $-$ 1 $\rightarrow$ LUMO) & 5.6 & 5.26 & 0.014 \\
& 2 & (HOMO $\rightarrow$ LUMO) & 16.82 & 16.43 & 0.014 \\

\hline
{${\mathrm{V}_{\mathrm{B}}^{-1}}$} 
& 2 & (HOMO $ \rightarrow$ LUMO) & 3.03 & 2.54 & 0.008 \\
& 2 & (HOMO $-$ 3 $\rightarrow$ LUMO) & 10.26 & 10.31 & 0.014 \\

\hline
\end{tabular}
\caption{Total HR Factors obtained from real excited state forces $\vec{F}^{\text{e}}$ and COHP-estimated forces $\vec{F}^{\text{COHP}}$ using full-phonon frequencies}
\label{table:HR}
\end{table*}

\begin{figure}[h!]
    \centering
    \includegraphics[width=0.9\linewidth]{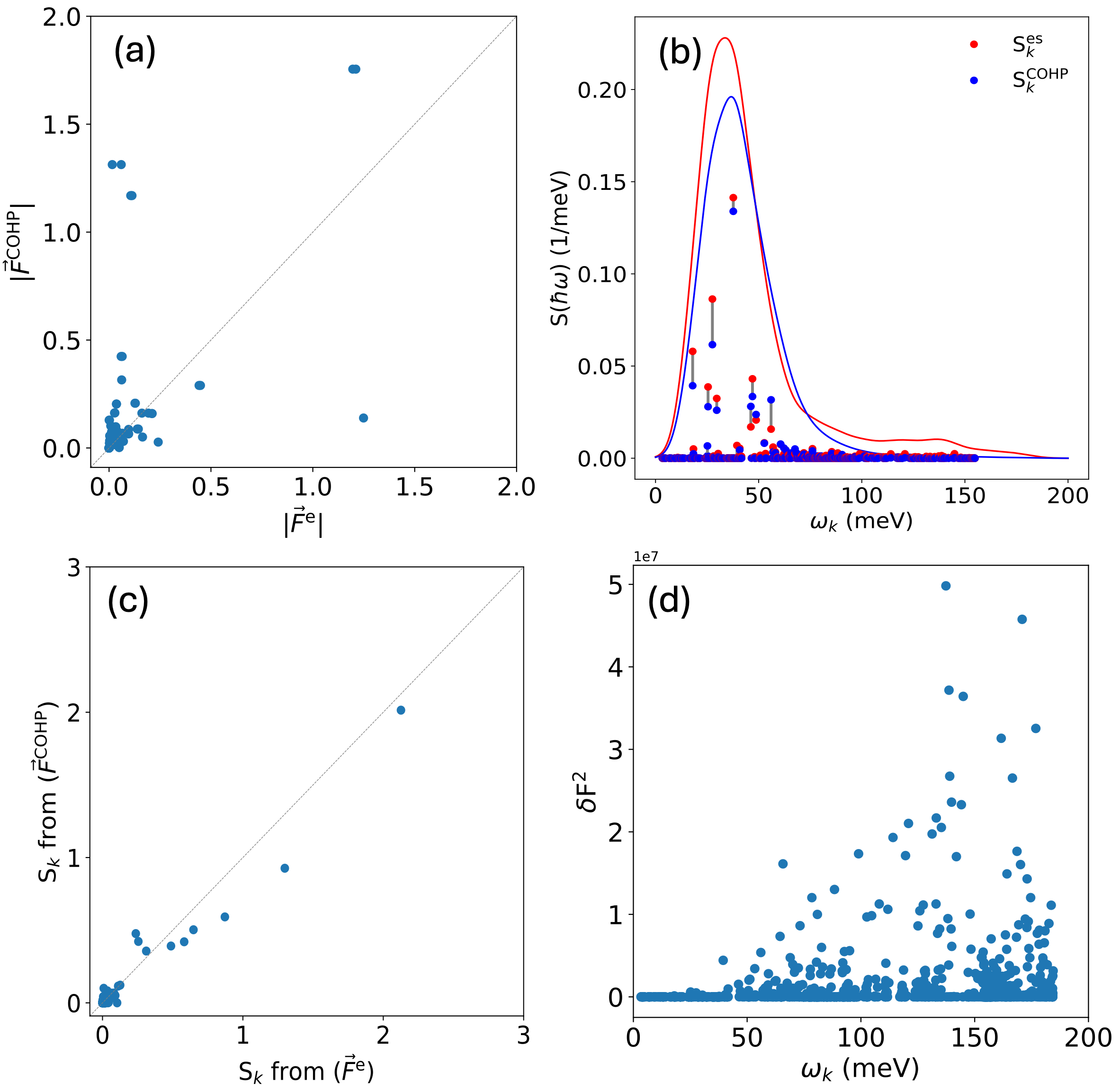}
    \caption[Outliers in estimated forces by COHP for an example transition of $\mathrm{V}_\mathrm{N}\mathrm{C}_\mathrm{B}$]{(a) Deviations in COHP-estimated forces for an example transition of $\mathrm{V}_\mathrm{N}\mathrm{C}_\mathrm{B}$. (b) Strong agreement in converged spectral functions obtained by estimated (blue) and real (red) partial HR factors. (c) Partial HR factors for the same transition, showing good agreement. (d) The errors of estimated forces projected on all phonon eigenvectors, showing large force differences projected onto more onto modes with higher frequencies.  }
    \label{fig:cohp-forces-outliers}
\end{figure}

\subsection{Excited-State Forces Comparison}
For some optical transitions examined in this paper we observe strong deviations in the estimated forces by COHP ($\vec{F}^{\text{COHP}}$). Fig.~\ref{fig:cohp-forces-outliers}a shows an example of these deviations in the (HOMO  $- 1 \rightarrow$ LUMO) transition for the $\mathrm{V}_\mathrm{N}\mathrm{C}_\mathrm{B}$ defect. However, in Fig.~\ref{fig:cohp-forces-outliers}c we show that this error in forces hardly propagates onto errors of similar magnitude in partial HR factors. The estimated partial HR factors in this transition have a small root mean square error of $\sim 0.02$ (see also Table \ref{table:HR}). The same error insensitivity can be observed in Fig.~\ref{fig:cohp-forces-outliers}b where the estimation error in the spectral function is also small. To determine the reason for this error insensitivity, in Fig.~\ref{fig:cohp-forces-outliers}d we consider the projection of errors in forces onto all phonon eigenvectors, plotted against phonon frequencies. It can be observed that the larger errors in COHP-estimated forces, if present as in this case, mainly project onto phonon eigenvectors with frequencies high enough ($>100$~meV here) that are beyond phonon modes that carry large $S_k$ dominating the spectral function lineshape (within 100~meV here). We expect that the force errors (when present) projecting mainly to high-frequency modes with negligible $S_k$ to apply more generally because the high-frequency limits of spectral functions are known to decay exponentially \cite{Norambuena2016, Leggett10987}, similar to what is shown in Fig.~\ref{fig:cohp-forces-outliers}b for the $\mathrm{V}_\mathrm{N}\mathrm{C}_\mathrm{B}$ case.

\end{document}